# Sub-second and ppm-level Optical Sensing of Hydrogen Using Templated Control of Nano-hydride Geometry and Composition

Hoang Mai Luong*[1], Minh Thien Pham[1], Tyler Guin[2], Richa Pokharel Madhogaria[3], Manh-Huong Phan[3], George Keefe Larsen*[2], and Tho Duc Nguyen*[1]

[1]*Department of Physics and Astronomy, University of Georgia, Athens, Georgia 30602, USA.*

[2]*National Security Directorate, Savannah River National Laboratory, Aiken, South Carolina 29808, USA.*

[3]*Department of Physics, University of South Florida, Tampa, Florida 33620, USA.*

*E-mail: hoanglm@uga.edu, george.larsen@srnl.doe.gov, ngtho@uga.edu

## Abstract

The use of hydrogen as a clean and renewable alternative to fossil fuels requires a suite of flammability mitigating technologies, particularly robust sensors for hydrogen leak detection and concentration monitoring. To this end, we have developed a class of lightweight optical hydrogen sensors based on a metasurface of Pd nano-patchy particle arrays, which fulfills the increasing requirements of a safe hydrogen fuel sensing system with no risk of sparking. The structure of the optical sensor is readily nano-engineered to yield extraordinarily rapid response to hydrogen gas (<3 s at 1 mbar $H_2$) with a high degree of accuracy (<5%). By incorporating 20% Ag, Au or Co, the sensing performances of the Pd-alloy sensor are significantly enhanced, especially for the $Pd_{80}Co_{20}$ sensor whose optical response time at 1 mbar of $H_2$ is just ~0.85 s, while preserving the excellent accuracy (<2.5%), limit of detection (2.5 ppm), and robustness against aging, temperature, and interfering gases. The superior performance of our sensor places it among the fastest and most sensitive optical hydrogen sensors.

**Main**

Hydrogen fuel is a key energy carrier of the future, and it is the most practical alternative to fossil fuel-based chemical storage, with a high theoretical energy density and universality of sourcing.[1] However, significant challenges remain with respect to the safe deployment of hydrogen fuel sources and therefore its widespread adoption.[2] For hydrogen leakage detection and concentration controls, it is essential that hydrogen sensors have good stability, high sensitivity, rapid response time, and most importantly be "spark-free".[3,4] High performance hydrogen sensors are, however, not only of importance in future hydrogen economy but also the chemical industry,[5] food industry,[6,7] medical applications,[8] nuclear reactors,[9] and environment pollution control.[10]

Numerous optical hydrogen sensors based on hydride-forming metal plasmonic nanostructure have been explored.[11,12] Pd is the most common hydriding metal for sensor applications due to its rapid response, room temperature reversibility, and relative inertness.[13] However, pure Pd nanoparticles suffer from the coexistence of an α-β mixed phase region, inducing hysteresis and hence non-specific readout and limited reaction rate.[12,14,15] It is possible to minimize this hysteresis and boost the reaction kinetics through the incorporation of alloying metals, such as Co, Au, or Ag.[11,12,16] It has been believed that the enhanced reaction kinetics in the alloying metal hydrides is associated with the reduction of the enthalpy of formation due to the reduced abrupt volume expansion occurring in smaller mixed phase regions, resulting in a reduction of energy barrier for hydride formation and dissociation, and the improved diffusion rate upon (de)hydrogenation.[12,15,17-20] Synergistically with material design, various sensing nanostructures have been engineered to minimize the volume-to-surface ratio (VSR) of the sensor, a critical condition for achieving fast response time and low hysteresis.[12] These structures include nanostripes,[21] nanoholes,[22,23] lattices,[23] nanobipyramids,[24] hemispherical caps,[23] nanowires, mesowires,[25] and chiral helices.[26,27] Moving beyond metals, other materials such a polymers can be incorporated, as Nugroho *et al.*[11] recently demonstrated, significantly boosting the sensitivity and response time of a Pd-based nanosensor. Along with downsizing the active material layer, the benchmark response time of 1 s at 1 mbar $H_2$ and 30 °C has been achieved,[11] however optical contrast was sacrificed. As a result, sub-second response time and ppm limit of detection (LOD) have not been achieved in a single sensor (Supplementary Table 1).

In this work, we demonstrate a compact optical hydrogen sensing platform with the fastest response reported to date and sub-10-ppm LOD. The sensor is comprised of Pd and Pd-alloy nano-patchy (NP) arrays with a simple optical transmission intensity readout. These hexagonally-packed nano-arrays are generated by facile single metal glancing angle deposition (GLAD) on polystyrene (PS) nanosphere monolayers. This fabrication process requires just one deposition step with no post-processing required, simplifying scale-up and reducing costs. The tunable film thickness, patchy diameter and shape (hemispheres or donuts), and therefore VSR can be readily controlled, resulting in tunable and rapid response rate. By alloying with cobalt, the response time of the metasurfaces was reduced below 0.85 s from 1-100 mbar of $H_2$ partial pressure, surpassing the strictest requirements for $H_2$ sensing while preserving excellent accuracy (<2.5% full scale) and 2.5-ppm LOD. In a broader perspective, our work illustrates evolution in hydrogen gas sensor technologies through rational topological design and targeted integration of non-traditional materials, such as polymers and active alloying elements. These concepts are universal for promoting strong interactions between gas and materials and may be generally applied to advance sensor and catalyst development, among others.

**Pd NP based optical hydrogen sensors**

The fabrication scheme and nanoarchitecture of an optical hydrogen sensor based on a hexagonal array of Pd hemispherical NPs are depicted in Fig. 1a (see Methods and Supplementary Figs. 1–5). A vapor incident angle of θ ≥ 50° was chosen to ensure that the film would not be deposited directly onto the underlying glass substrate (Supplementary Fig. 2).[23] The designed structure consists of NP arrays on top of hexagonal closed-packed nanosphere monolayer, which is confirmed by SEM and energy-dispersive spectroscopy (EDS) elemental mapping (see Fig. 1b–d and Supplementary Figs. 6–8). Note that NP samples with a specific deposited thickness will now be referred to as $NP_t^\theta$, where θ (°) and $t$ (nm) represent the incident angle of the metal vapor and the nominal thickness of the deposited film, respectively.

The morphological transition of Pd $NP_{t_{Pd}}^\theta$ with an increasing $t_{Pd}$ was observed using ultra-high-resolution SEM (SU-9000, Hitachi), as revealed in Fig. 1d. The morphology of $NP_{1.5}^{50}$ contains many sub-10-nm granules, and these granules fully cover the top surface of the polystyrene nanosphere. The size of the granules grows when $t_{Pd}$ = 2 nm, and coalescence via a neck (or bridge) connection between neighboring clusters[28] can be noticed at the thickness of $t_{Pd}$ = 3.5 nm. A continuous film is formed at a thickness of 5 nm, as these bridge connections are successively grown. Once the continuous film is shaped at a thickness of 5-nm, another distinct coalescent mechanism is observed: the size of the clusters increases as the thickness increases, associated with the reduction of the cluster density (see Supplementary Fig. 8 for a detail analysis). The influences of the film morphology on the thermodynamic response of NP samples will be discussed in the sections below.

The transmission magnitude of the Pd metasurfaces monotonically increases and approaches that of a bare PS nanosphere monolayer with decreasing $t_{Pd}$ (Fig. 2a). Effects from the PS monolayer are observed in the transmission spectra of all films, such as the optical band gap located at wavelength $\lambda \approx 618$ nm and other transmission maxima/minima at shorter wavelengths due to interference effects.[29] Upon hydrogenation, the optical transmission ($T_{P_{H_2}}$) of all NPs increases with a strong dependence on wavelength as shown in Fig. 2b, where $\Delta T(\%) = T_{P_{H_2}} - T_0$ is the difference of transmission intensity when $P_{H_2}$ = 1000 and 0 mbar. The maximum optical change is much greater than that of the control Pd thin film sample (Supplementary Section 3). Note that the optical transmission of a stand-alone PS nanosphere monolayer does not change upon its exposure to hydrogen (Supplementary Section 3). The increasing trend of optical transparency upon hydrogenation has been also observed in Pd thin films, although this increase is independent of wavelength in the visible-NIR region.[30,31]

Optical hydrogen sorption isotherm of $NP_{t_{Pd}}^{50}$ samples are shown in Fig. 2c, where the change of transmission intensity $\Delta T(\%)$ versus $P_{H_2}$ is calculated at the spectrum maxima to achieve the highest signal-to-noise ratio (SNR). We note that the plateau pressure could be measured at any wavelength within the visible region. Several important dependencies can be observed from the optical isotherms. Primarily, $\Delta T$ is generally reduced as $t_{Pd}$ decreases due to lower overall light-patch-interaction signal. The mixed region hysteresis is reduced, and the slope of the plateau pressure increases with decreasing $t_{Pd}$. The size dependences of the plateau slope and hysteresis have been previously observed in Pd nanoparticles with a narrow size distribution, but we cannot discount the effect of size distribution on the slope for $NP_{t_{Pd}}^{50}$ samples (Supplementary Figs. S8–9).[32]

Figure 2d displays the plateau pressure of absorption isotherm, desorption isotherm, and hysteresis ($P_{Abs}$, $P_{Des}$, and $\ln(P_{Abs}/P_{Des})$, respectively) *versus* $t_{Pd}$ (the extraction method is described in Supplementary Section 4). We observe a critical thickness ($t_{Pd} \approx 5$ nm) above which the nanoparticles behave qualitatively differently than for thicknesses below. It is worthwhile to note that this critical thickness corresponds well with the expected transition from a separated island-like morphology to a percolating film morphology. Above this critical thickness, $P_{Abs}$ and $P_{Des}$ are relatively independent of $t_{Pd}$ and $\ln(P_{Abs}/P_{Des}) \approx 0.62$, which is the value predicted for bulk Pd undergoing a coherent phase transition.[33] Below the critical thickness, both $P_{Abs}$ and $P_{Des}$ sharply increase, while $\ln(P_{Abs}/P_{Des})$ sharply decreases. This change in $P_{Abs}$ and $P_{Des}$ is due to the increasing effects of higher energy subsurface adsorption sites relative to bulk sites.[32,34] The significant decrease in hysteresis for shrinking particle size at a constant temperature of Pd material, $T$, has been widely observed and is understood as a decrease in critical temperature, $T_c$, with decreasing the particle size (Supplementary Fig. 11).[35] In general, when $T$ is close to $T_c$, the hysteresis behavior is strongly suppressed. See Supplementary Subsection 2.3 for more in depth analyses of $P_{Abs}$, $P_{Des}$, and $\ln(P_{Abs}/P_{Des})$ as functions of $t_{Pd}$.

**Performance of Pd metasurfaces**

The hysteresis of an optical sensor, and therefore its insensitivity to measurement history, can be calculated in terms of "sensor accuracy", as proposed by Wadell et al. (Supplementary Section 5).[36] Figure 2e summarizes the sensor accuracy of $NP^{50}_{t_{Pd}}$ with varying $t_{Pd}$, from $10^1$ to $10^6$ μbar $P_{H_2}$. The $P_{Hys}$ ($= P_{Abs} - P_{Des}$) reported here (~7.5 mbar in $NP^{50}_{15}$, ~4 mbar in $NP^{50}_{2}$) is significantly smaller than $P_{Hys}$ reported for any other optical based nanostructure/thin film sensing platforms made from pure Pd (typically $P_{Hys} > 20$ mbar).[11,36] Therefore, these NP-based sensors offer extremely high sensor accuracy (<10 % at $t_{Pd} \leq 15$ nm and <5 % at $t_{Pd} \leq 5$ nm), which is comparable to that of Pd-alloy based hydrogen sensors.[36,37]

The response time of the NP sensor is improved with decreasing $t_{Pd}$. The response times ($t_{90}$, the time required to reach 90% of the final equilibrium response) of $NP_{t_{Pd}}^{50}$ sensors with varying $t_{Pd}$ were measured when varying pressure pulses from 100 to 1 mbar $H_2$. The $t_{90}$ generally decreases with increasing $P_{H_2}$, though there is a peak at $P_{Abs}$ (Fig. 2f). The peak in the $t_{90}$ versus $P_{H_2}$ relation for $P_{Abs}$ has been reported in some Pd-based hydrogen sensor,[23,38,39] however, the origin of this observation has not been described. We hypothesize that the peak is associated to the abrupt change in hydride volume during the $\alpha - \beta$ phase transition. Such abrupt change or sudden entropy change occurs only if the hydride system thermodynamically overcomes the large strain-induced energy barrier upon hydrogenation, which requires longer time to reach the equilibrium. The strain-induced energy barrier is relatively strong in pure Pd based sensors, that results in the large $t_{90}$ peak in this pressure regime. It is clear in Fig. 2f that such barrier is larger with a larger grain size and film thickness,[12] and interestingly, the barrier becomes much less significant as the structure transforms from film-like to nanoparticle island-like for $t_{Pd} \leq 2$ nm. The $t_{90}$ of $NP_2^{50}$ is <3 s over the 1-100 mbar range, which is nearly two orders of magnitude smaller than that of $NP_{15}^{50}$ (the longest $t_{90}$ is ~120 s). Crucially, the $t_{90}$ of $NP_2^{50}$ and $NP_5^{50}$ samples at 40 mbar (the lower flammability limit of hydrogen) is ~1 s, which is shorter than that of previously reported pure Pd-based optical sensors (~7.5 s,[11] or typically >20 s),[38] and comparable to that of Pd-alloy based sensors.[36,37] In general, the response time as a function of $t_{Pd}$ follows a power-law scaling with respect to the VSR ($VSR^z$), with an exponent of $z \approx 2.3$, which agrees well with a diffusion limited metal hydride system (Supplementary Subsection 2.4).[40,41]

The flexibility of the GLAD technique allows the realization of different nanopatterns of patchy array sample by adjusting the polar and azimuthal angle of deposition onto the packed nanosphere monolayers (Supplementary Figs. 2–3).[42] Here, we utilize θ = 70°, 80°, and 85° with constant azimuthal rotation (fixed $t_{Pd}$ = 15 nm) to engineer the morphology and VSR of NPs. Figure 3a displays SEM micrographs of $NP_{15}^{50}$, $NP_{15}^{70}$, $NP_{15}^{80}$, and $NP_{15}^{85}$ samples, which confirm the predicted surface morphologies. Figure 3b displays the optical hydrogen sorption isotherm of these samples at maximum $\Delta T(\lambda)$. The amount of material deposited decreases as $\theta$ increases, and correspondingly, the samples are generally more transparent and $\Delta T$ changes less upon hydrogenation (Fig. 3b). Apart from the transmission magnitude change, we observe a narrowing of hysteresis in the mixed phase region and an increasing slope of plateau as $\theta$ increases. The phase transition behavior of NP is further shown in Supplementary Section 7.

As mentioned, a key advantage of this nano-fabrication method is the straightforward engineering of the VSR, which can decrease sensor response times.[11] By increasing $\theta$, one can efficiently reduce VSR since the patchy pattern gradually transforms from hemisphere-like to donut-like structures. A strong $\theta$-dependent hydriding kinetics of $NP_{15}^{\theta}$ can be observed in Fig. 3c. For example, the $t_{90}$ of a $NP_{15}^{85}$ is <10 s over pressures 1-100 mbar, which is nearly an order of magnitude smaller than these of a $NP_{15}^{50}$. More significantly, we observe a general linear relation between VSR and $t_{90}$ at different $P_{H_2}$ of NP sample, regardless of vapor incident angle $\theta$ and deposited thickness $t_{Pd}$ (Fig. 3d). Note that the estimation of VSR is based on simulated morphologies,[43] (Supplementary Section 1) and there may be some deviations from the achieved parameters due to non-idealities, surface roughness, and island morphologies. However, the general trend of creating faster absorption kinetics was observed when the VSR and/or $t_{Pd}$ was reduced. Additionally, we highlight that the VSR may be further optimized to achieve even shorter response time by other nanoparticle forms, such as nano-fan,[42] which can be created by modifying depositions onto packed nanosphere monolayers with different polar and azimuthal angles.

**Pd-composite NP optical hydrogen sensors**

To further optimize the accuracy and response time of the nano-sensor, we utilized Pd-based composites with alloying elements (Ag, Au, or Co) to modify the hydriding properties of Pd. For example, PdAg and PdAu alloys display lower plateau pressures than pure Pd,[11,23,36,44] while PdCo alloys display much higher plateau pressures.[45,46] $Pd_{80}Ag_{20}$, $Pd_{80}Au_{20}$, and $Pd_{80}Co_{20}$ composite $NP_5^{50}$ and $NP_{15}^{50}$ samples were fabricated by employing the electron beam co-evaporating method. It is worth noting that the $Pd_{80}Au_{20}$ sensors provides a useful direct comparison to the major works on PdAu hydride systems.[16,36,47-49] Utilizing EDS elemental mapping, we confirm the composition and uniform element distribution of Pd and Ag/Au/Co in the composite NP samples (Fig. 4a).

The hydrogen sorption characteristics of the $Pd_{80}Ag_{20}$ alloy NP system are examined as previously (Supplementary Fig. 19). The plateau of $Pd_{80}Ag_{20}$ NP is shifted downward to lower pressure ($P_{Abs} \approx 1$ mbar) and shows an increasing slope in comparison to that of the Pd NP (Supplementary Fig. 19c). The hysteresis is very narrow ($P_{Hys}$ are <0.1 mbar and <0.05 mbar in $Pd_{80}Ag_{20}$ $NP_{15}^{50}$ and $NP_5^{50}$, respectively), and the optical isotherm could be considered "hysteresis-free" (Fig. 4b). These behaviors of $Pd_{80}Ag_{20}$ NP system are in agreement with bulk PdAg alloy,[44] and are similar to behaviors of PdAu alloy nanoparticles and thin films with higher Au content.[36,37,50] We note that the $P_{Abs}$ of PdAg composite NP ($P_{Abs} \approx 1$ mbar) is significantly smaller than those of the previously employed systems (typically $P_{Abs} \geq 10$ mbar).[36,37,50] In addition, the absorption kinetics are noticeably accelerated as the $t_{90}$ of $Pd_{80}Ag_{20}$ sensors are significantly faster than that of Pd sensors with the same deposited thickness (Fig. 4d). The origin of this acceleration could be the reduction of the apparent activation energy for hydrogen sorption due to combination with Ag,[51] change in morphology to well-separated islands (Supplementary Fig. S10), and due to a decrease in abrupt volume expansion occurring in a smaller mixed phase region, as observed in the narrow hysteresis[52]; both leading to a significant increment of hydrogen permeability in $Pd_{80}Ag_{20}$.[53] The pressure dependent response time of $Pd_{80}Ag_{20}$ sensors follows the power-law[11] as confirmed by an near-linear dependence between $\log(t_{90})$ versus $\log(P_{H_2})$.

In comparison to PdAg system, PdAu samples with the same thickness exhibits a similar spectra shape and transmission magnitude (Supplementary Fig. 19a). Upon the hydrogenation, $Pd_{80}Au_{20}$ $NP_5^{50}$ and $NP_{15}^{50}$ samples behave similar to $Pd_{80}Ag_{20}$ $NP_5^{50}$ and $NP_{15}^{50}$, with comparable optical transmission changes ($\Delta T(\lambda)$) and similar spectra shape (Supplementary Fig. 19b). However, the plateau pressures found in $\Delta T(\lambda)$ optical isotherms of PdAu samples are at higher pressure regimes ($P_{Abs} \approx 7$ mbar) than these of PdAg systems ($P_{Abs} \approx 1$ mbar) (Fig. 4b). We note that the $P_{Abs}$ we found in PdAg NPs are slightly smaller than that of PdAu nanoparticles with a similar composition ($P_{Abs} \geq 10$ mbar).[36,37,50] In addition, a noticeable hysteresis can be seen in $\Delta T(\lambda)$ isotherms of $Pd_{80}Au_{20}$ $NP_{15}^{50}$ ($P_{Hys} \approx 1.1$ mbar) and $Pd_{80}Au_{20}$ $NP_5^{50}$ ($P_{Hys} \approx 0.7$ mbar), which are wider than these of PdAg samples ($P_{Hys}$ are <0.1 mbar). However, the sensor accuracies calculated over the pressure range of $10^1$ to $10^6$ μbar are still below 5% level (Fig. 4c). At a given pressure $P_{H_2} > P_{Abs} \approx 7$ mbar, response time ($t_{90}$) of a PdAu sensor is slightly higher than that of a PdAg sensor with the same thickness, which follows the power-law with a similar slope (Fig. 4d). Interestingly, we observe a kink at $P_{Abs} \approx 7$ mbar, below which $t_{90}$ of PdAu still follows the power law, however, with a noticeable smaller slope. The $t_{90}$ of $Pd_{80}Au_{20}$ $NP_{15}^{50}$ and $NP_5^{50}$ samples at 1 mbar are 13 s and 5 s, respectively, which are much shorter than these of PdAg sensors (45 s and 9 s for $NP_{15}^{50}$ and $NP_5^{50}$ samples, respectively). Since the kink is observed in the middle of the plateau, which is in a similar position as that in the pure Pd sensor case (Fig. 2f), we hypothesize that the strain-induced energy barrier in the $Pd_{80}Au_{20}$ is still significant at $P_{Abs} \approx 7$ mbar and is the cause of this observation. We note that the differences in $t_{90}$ between the PdAu and PdAg samples with the same $t$ and $P_{H_2}$ might come from several factors, such as differences in morphologies (see Supplementary Figs. 9–10), and hydrogen permeability.[54]

Several advantages are achieved through the utilization of a PdAg and PdAu composite instead of pure Pd in a representative $NP_{15}^{50}$ sensor, which are listed as (**1**)-(**4**). (**1**) The "hysteresis-free" optical isotherm removes undesirable uncertainty, generating an excellent sensor accuracy of <5 % over the pressure range of $10^1$ to $10^6$ μbar (Fig. 4c). (**2**) The slope of absorption isotherm of PdAg and PdAu samples in the $\alpha$-phase ($P_{H_2} < 0.5$ mbar) shows a 3- and 1.8-times enhancement in sensitivity as compared to pure Pd sensors, respectively (Supplementary Fig. 19c), which benefits trace level detection. (**3**) Particularly for PdAg system, the plateau shifts to very low pressures ($P_{Abs} \approx 1$ mbar), which enhances the sensitivity by an order of magnitude throughout the pressure range of 0.5 mbar to 10 mbar (~0.05 – 1 vol%). This is essential for early leak detection (4 vol% is the lower flammability limit of hydrogen). (**4**) The absorption kinetics are accelerated (Fig. 4d), which yields an order of magnitude $t_{90}$ faster than that of pure Pd at the plateau pressure.

While the advantages (**1**)-(**3**) achieved by $Pd_{80}Ag_{20}$ and $Pd_{80}Au_{20}$ $NP_{15}^{50}$ sensors are comparable to or surpasses the performances of up-to-date optical composite-based nano-sensors,[11,16,36,37] the accelerated response time (~45 s and ~13 s at $P_{H_2} = 1$ mbar, respectively) in (**4**) is still too slow to fulfill the most rigorous requirement of a hydrogen sensor for automotive applications.[3] Engineering the VSR by downsizing NP significantly reduces this gap, as observed in the $Pd_{80}Ag_{20}$ and $Pd_{80}Au_{20}$ $NP_5^{50}$ sample. While the shape of optical sorption isotherm is relatively unchanged and the sensor accuracy of <5 % is preserved upon thickness reduction from 15 nm to 5 nm (Fig. 4b,c), the $t_{90}$ of $Pd_{80}Ag_{20}$ and $Pd_{80}Au_{20}$ $NP_5^{50}$ is reduced by a factor of ~5 and ~2.6 over a pressure range 1-100 mbar and is only ~9 s and ~5 s at $P_{H_2}$ = 1 mbar, respectively. The slopes of the linear fits in ($\log(t_{90})$ *versus* $\log(P_{H_2})$) plot of $NP_{15}^{50}$ and $NP_5^{50}$ samples are equivalent for both PdAg and PdAu systems (Fig. 4d), which implies that these two samples have similar sorption behaviors and that the improvement of the response time is primarily derived from the reduction of VSR.[11] Hence, in order to achieve the $t_{90} \leq 1$ s at $P_{H_2}$ = 1 mbar, the deposited thickness of $Pd_{80}Ag_{20}$ and and $Pd_{80}Au_{20}$ should be <1 nm (see Supplementary Section 1 and Supplementary Fig. 13). A minimum deposited thickness of $t$ = 1.5 nm is required for overcoming the bead's roughness and obtaining a detectable optical signal contrast in the pure Pd $NP_t^{50}$ at $P_{H_2}$ = 1 mbar. This means that it would be impossible to attain a $Pd_{80}Ag_{20}$ $NP_{15}^{50}$ sensor with $t_{90} \leq 1$ s at $P_{H_2}$ = 1 mbar simply by reducing $t$.

In order to reduce the response time further, the sorption behavior of the NP sensors was modified through the incorporation of Co ($Pd_{80}Co_{20}$), analogous to $Pd_{80}Ag_{20}$ and $Pd_{80}Au_{20}$. A PdCo alloy improves the kinetics of hydrogenation over a PdAg and PdAu alloy by (1) remaining in the α-phase over the pressure region of interest, which removes the kinetic steps of the α- to β-phase transition and subsequent hydrogen atomic diffusion through the β-phase;[55] and (2) cobalt offers a greater metal-hydrogen bond strength than silver, which could facilitate dissociative chemisorption of hydrogen.[56] It is also worth noting that the comparison between the PdAu and PdCo NP sensors allows isolation of the alloying element effect since their morphologies are very similar (Supplementary Fig. S10). In the $Pd_{80}Co_{20}$ sensors, the overall shape of $\Delta T(\lambda)$ is similar to those of the Pd, $Pd_{80}Ag_{20}$, and $Pd_{80}Au_{20}$ sensors. However, the magnitude of $\Delta T$ is significantly smaller as PdCo alloy has lower H-solubility due to lattice contraction (Supplementary Fig. 19).[45]

Figure 4b shows the $\Delta T$–$P_{H_2}$ isotherm of $Pd_{80}Co_{20}$ $NP_5^{50}$ extracted at the maximum $\Delta T(\lambda)$. The plateau pressure is shifted significantly upward, and consequently, no hysteresis in the sorption isotherm is observed in the measured pressure range, which is consistent with the previous observation for bulk PdCo alloys and similar to the behaviors of the PdCu alloy nano-sensors.[37,45,46] The "hysteresis-free" characteristic is reflected in the very high sensor accuracy (<2.5%) (Fig. 4c). Additionally, remarkably accelerated absorption in the PdCo composite is observed (Fig. 4d). Two notable results are emphasized: (**1**) the $t_{90}$ at $P_{H_2}$ = 1 mbar is 0.85 s, which satisfies the most stringent requirement of a hydrogen sensor (<1 s over pressures 1-100 mbar), matching the fastest optical hydrogen nano-sensor under similar conditions,[11] and (**2**) the $t_{90}$ at $P_{H_2}$ = 40 mbar (~ 4 vol %) is ~0.15 s (the resolution limit of our system is at 0.15 s), which is nearly twice as fast as those of previously reported optical hydrogen nano-sensors under similar conditions.[11,16,36,37] In addition, we highlight that (**i**) coating of a polymer thin film and (**ii**) optimizing the VSR may further improve this ultra-fast response time.[11]

One shortcoming of downsizing the active material layer is that the optical contrast upon (de)hydrogenation decreases significantly, which makes it challenging to achieve ultra-fast response time and ultra-low LOD in a single nano-sensor. The unique design of the NP sensor allows for a very high surface coverage (>90 %, Supplementary Fig. 17) and results in sizable optical contrast even at very low concentration of hydrogen. In Fig. 5a, the detection capability of the PdCo $NP_5^{50}$ hydrogen sensor is demonstrated in step-wise pressure pulses of pure $H_2$ from 5000 to 11 μbar at a 1.25 Hz sampling rate (11 μbar is the lower pressure limit of our measurement setup). Clearly, PdCo NP$_5^{50}$ can resolve the lowest pulse and potentially provide detection at even lower hydrogen pressures. The PdCo NP$_5^{50}$ sensor is further tested in flow-mode with different hydrogen concentration ($C_{H_2}$) from 100 to 1 ppm in nitrogen (Fig. 5b), and 3-cycles of each $C_{H_2}$ from 40000 to 10 ppm in nitrogen (Supplementary Fig. 20). The results show reproducible responses that are distinct from the background signal at $C_{H_2}$ as low as 10 ppm, with the possible experimental LOD as low as 2.5 ppm (Supplementary Figs. 26–27). We note that results for similar tests with PdAg NP$_{15}^{50}$ and NP$_5^{50}$ sensors can be seen in Supplementary Fig. 22. Further sensing test with synthetic air as a carrier gas shows that the PdCo NP$_5^{50}$ sensor exhibits a smaller response at a given $C_{H_2}$, in comparison to the responses of the sensor in vacuum/inert carrier gas (Supplementary Fig. 21). This is due to the competing processes of the hydrogen oxidation reaction and the adsorption sites are blocked by O.[37,38] The $\Delta T(\lambda)/T$ amplitude of the sensors is extracted and summarized in Fig. 5c, which demonstrates approximately equivalent optical responses in vacuum-mode sensing and flow-mode sensing with nitrogen carrier gas, and smaller responses in flow-mode sensing with synthetic air carrier gas. By defining LOD = 3$\sigma$, where $\sigma$ is the noise of the acquired signal (which is 0.0042 %, see Supplementary Section S9), we achieved the LOD of 2.5 ppm in nitrogen and 10 ppm in air. These experimental performances (LOD of 2.5 ppm and $t_{90} \approx 0.85$ s at 1 mbar, in a single sensor) places our PdCo NP$_5^{50}$ system among the fastest and most sensitive optical hydrogen sensors (see Supplementary Section 12).[11,39]

In order to assess the practical implementation of these hydrogen sensors, the influence of temperature, moisture, and interfering gases (e.g., CO, $CO_2$, $CH_4$) on the sensor performances were investigated. For the PdCo NP$_5^{50}$ sensor, we found that neither 5% $CO_2$ nor 5% $CH_4$ in the 2% $H_2$ feed gas affects to the sensor signal (Fig. 6a,b). In addition, PdCo NP$_5^{50}$ sensor shows a good tolerance toward the temperature, as the signal amplitude stays at ~80% of the reference signal, up to 315 K (Fig. 6c,e). However, a moderate sensor poisoning occurs upon the exposure to the 0.2 % CO in the feed gas, as the signal amplitude decreases to ~60% of the reference signal (Fig. 6a,b). We note that the sensor signal can be fully recovered by multiple (de)hydrogenation cycles (Supplementary Fig. 29). In addition, we observed a ~30% drop in signal amplitude when the synthetic gas carrier air has a relative humidity of 40% (Fig. 6d,f). This desensitization is understandable, as the water molecules are in contact and dissociatively adsorbed by the Pd-surface, which induces the consumption of $H_2$ and forms the gas phase water.[57] In term of sensor stability upon cycling, the sensor signal is reproducible upon several hundred of (de)hydrogenation cycles without any sign of degradation (Supplementary Fig. 27a,b). However, the sensor signal slowly changes upon long-term storing in the ambient conditions, as the sensor does not completely immune from trace gases present in air, such as CO and moisture. We observed a reduction of sensor signal, as well as decay performances in response time (still, the $t_{90}$ and LOD in nitrogen background of PdCo NP$_5^{50}$ are <2.5 s (at 1 mbar) and <10 ppm, respectively, over a >10-month period and >400 (de)hydrogenation cycles) (Supplementary Fig. 27c-e). Nevertheless, a protective layer, such as a polymer-coating, would protect the sensor against humidity, CO, and other toxic gases and ensure long term reliability.

Inspired by previous works,[11,58,59] our preliminary results demonstrate that a simple polymer coating layer of polymethyl methacrylate (PMMA) achieved by spin-coating can provide excellent protection for PdCo NP$_5^{50}$ sensor against CO and humidity. The standard sensing characterizations of this PdCo NP$_5^{50}$/PMMA sensor can be found in Supplementary Fig. 30. In both deactivation tests with poisonous gases ($CH_4$, $CO_2$, CO, in Fig. 7a,b) and humidified carrier gases (RH = 40%, in Fig. 7c,d), the absolute response of the sensor remains within the ± 20% deviation limit, which satisfy the standard requirement for hydrogen sensors.[60] We note that in this case, the PMMA polymer layer protects the Pd-adsorption domain from direct contact with water molecules, which reduces the $H_2O$-induced $H_2$ consumption and prevents the desensitization.[57] The water vapor condensing on the polymer coating does have an effect by impeding $H_2$ reaching to adsorption domain, and slightly increasing the response/releasing times (Fig. 7c).[61,62] Notably, the polymer coating layer significantly mitigates the degradation of sensing performance when this sensor operates in a practical condition (Supplementary Fig. 28). This is an interesting improvement for this hydrogen sensing platform that warrants further study.

**Conclusions**

We have demonstrated a method to produce a novel class of rapid-response, highly sensitive, and accurate optical Pd-alloy hydrogen sensors through GLAD on PS-nanospheres. It is facile to tune these metasurface sensors through simple alloying, angle of deposition, or film thickness, which dictates the qualitative nature, quantitative metrics, and hysteresis of the response. By incorporating 20% Co, the sensor response time ($t_{90}$) at 1 mbar is less than 0.85 s, which is the fastest response ever reported at the critical $H_2$ concentration required for leak detection. Additionally, these sensors readily detect concentration as a low as 2.5 ppm in nitrogen and 10 ppm in air, maintain high accuracy due to the hysteresis-free operation, and exhibit robustness against aging, temperature, moisture, and disturbance gases. These sensors demonstrate a viable path forward to spark-proof optical sensors for hydrogen detection applications.

**Methods**

**Materials.** Polystyrene (PS) nanospheres (Polysciences Inc., $D = 500 \pm 10$ nm) and ethanol (Sigma-Aldrich, 98%) were used to create the nanosphere monolayers. Palladium (99.95%), silver (99.99%), and cobalt (99.95%) from Kurt. J Lesker Company were utilized for electron beam depositions. Deionized water (18 MΩ·cm) was used for all experiments.

**Sample fabrication.** Hexagonal close-packed nanosphere ($D = 500$ nm) monolayers on glass and Si substrates ($1 \times 1$ cm$^2$), which were prepared by an air/water interface method,[23,63-65] were used as a template for electron beam deposition. For pure Pd NP samples, the substrates were coated by Pd with a varied thickness of $t_{Pd}$, under a constant deposition rate of 0.05 nm/s, and the sample holder rotated azimuthally with a constant rotation rate of 30 rpm during deposition process. For Pd-Ag, Pd-Au or Pd-Co composite NP sample, materials were placed in two independent crucibles on two sides of the chamber, and the vapor incident angles to substrate normal were 10° and -10°, respectively. The deposition rates and thicknesses of Pd and Ag/Au/Co were monitored independently by two separated quartz crystal microbalances (QCM). By controlling the deposition rates of Pd and Ag/Au/Co, $Pd_{80}Ag_{20}$, $Pd_{80}Au_{20}$, and $Pd_{80}Co_{20}$ NPs were realized.

**Morphology and composition characterization.** Scanning electron microscopy (SEM) was performed with a Thermo Fisher Scientific (FEI) Teneo field emission scanning electron microscope (FESEM). Energy-dispersive spectroscopy (EDS) elemental mapping was performed with 150 mm Oxford XMaxN detector. Ultra-high-resolution SEM was performed with a SU-9000, Hitachi (with resolution of 0.4 nm at 30 kV).

**Hydrogen sensing measurement.** All optical isotherm, LOD, response/release time measurements are performed in a home-made vacuum chamber with two quartz windows.[23] The hydrogen pressure was monitored by three independent pressure transducers with different which cover the pressure range of $10^{-6}$ to 1.1 bar (two PX409-USBH, Omega and a Baratron, MKS). Optical transmission measurements were performed with an unpolarized collimated halogen lamp light source (HL-2000, Ocean Optics) and a spectrometer (USB4000-VIS-NIR-ES, Ocean Optics). The optical response/release time measurements were performed at 12.5 Hz sampling frequency (4 ms integration time with 10 averages). The LOD measurements were performed at 1.25 Hz sampling frequency (4 ms integration time with 100 averages), and $\Delta T/T$ responses were averaged over wavelength range of $\lambda = 500 - 660$ nm for the best SNR. For LOD measurements in flow mode, ultra-high purity hydrogen gas (Airgas) was diluted with ultra-high purity nitrogen gas (Airgas) or synthetic gas (Airgas) to targeted concentrations by commercial gas blenders (GB-103, MCQ Instruments). The gas flow rate was kept constant at 400 ml/min for all measurements. All experiments (except the temperature-dependent experiments) were performed at constant 25 °C.

**FDTD calculations.** FDTD calculations of Pd hydride NP samples were carried out using a commercial software (Lumerical FDTD Solutions).[66] The geometric parameters of Pd cap were obtained from MATLAB simulation. The mesh size of 2 nm × 2 nm × 2 nm was chosen. The refractive index of glass and PS was chosen to be 1.5 and 1.59, respectively, and the optical parameters of Pd and $PdH_x$ were extracted from Ref. [67].

**PMMA coating.** PMMA (Sigma Aldrich, 10 mg/ml dissolved in acetone by heating up the mix to 80 °C and cooling down to the room temperature, $M_w = 15\,000$) was spin-coated on sensors at 5000 r.p.m. for 120 s followed by a soft baking at 85 °C on a hotplate for 20 minutes. Using the same coating process on a clean glass substrate results in a ~50 nm PMMA film, as measured by an atomic force microscope (NX-10, Park Instrument).

**Acknowledgements**

The authors thank Prof. Yiping Zhao for his generosity in sharing his nano-fabrication tools with us. This work was supported by Savannah River National Laboratory's Laboratory Directed and Development program (SRNL is managed and operated by Savannah River Nuclear Solutions, LLC under contract no. DE-AC09-08SR22470). M.H.P. acknowledges support from the U.S. Department of Energy, Office of Basic Energy Sciences, Division of Materials Sciences and Engineering under Award No. DE-FG02-07ER46438.

**Conflict of Interest**

The authors declare no conflict of interest.

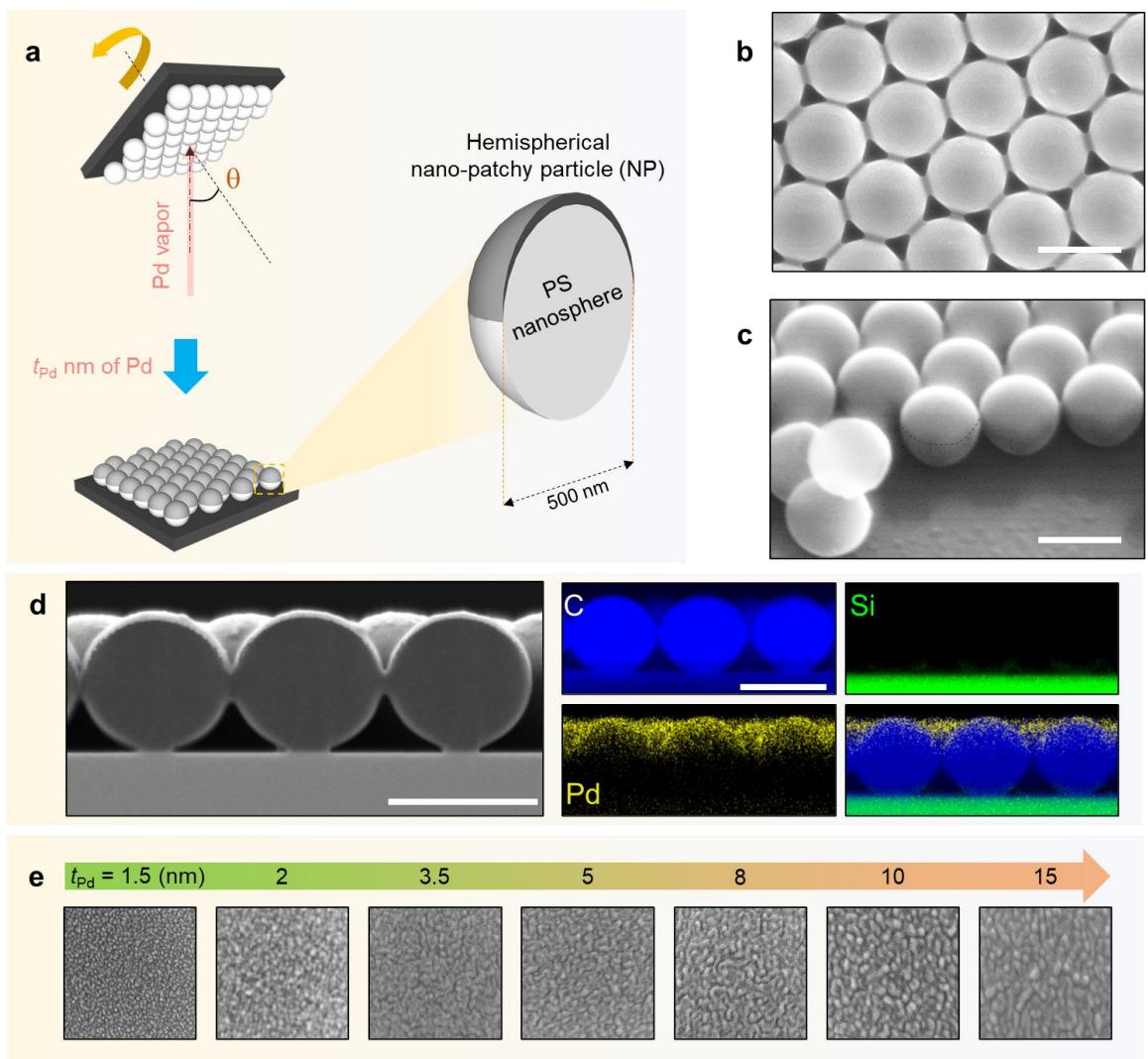

**Fig. 1 | Fabrication scheme and morphology characterization.** (a) Schematic of the fabrication process. (b) Top-view and (c) side-view scanning electron microscope (SEM) images of Pd $NP_{15}^{50}$. Scale bars correspond to 500 nm. (d) Cross-sectional SEM image and EDS elemental maps of $NP_5^{50}$ showing a metal cap formed on the top surface of a PS nanosphere. Scale bars correspond to 500 nm. (e) Magnified views (150 nm × 150 nm) of ultra-high resolution SEM micrographs (in Supplementary Fig. 8) showing morphology differences of Pd NP with different deposited thickness ($t_{Pd}$) of 1.5 to 15 nm.

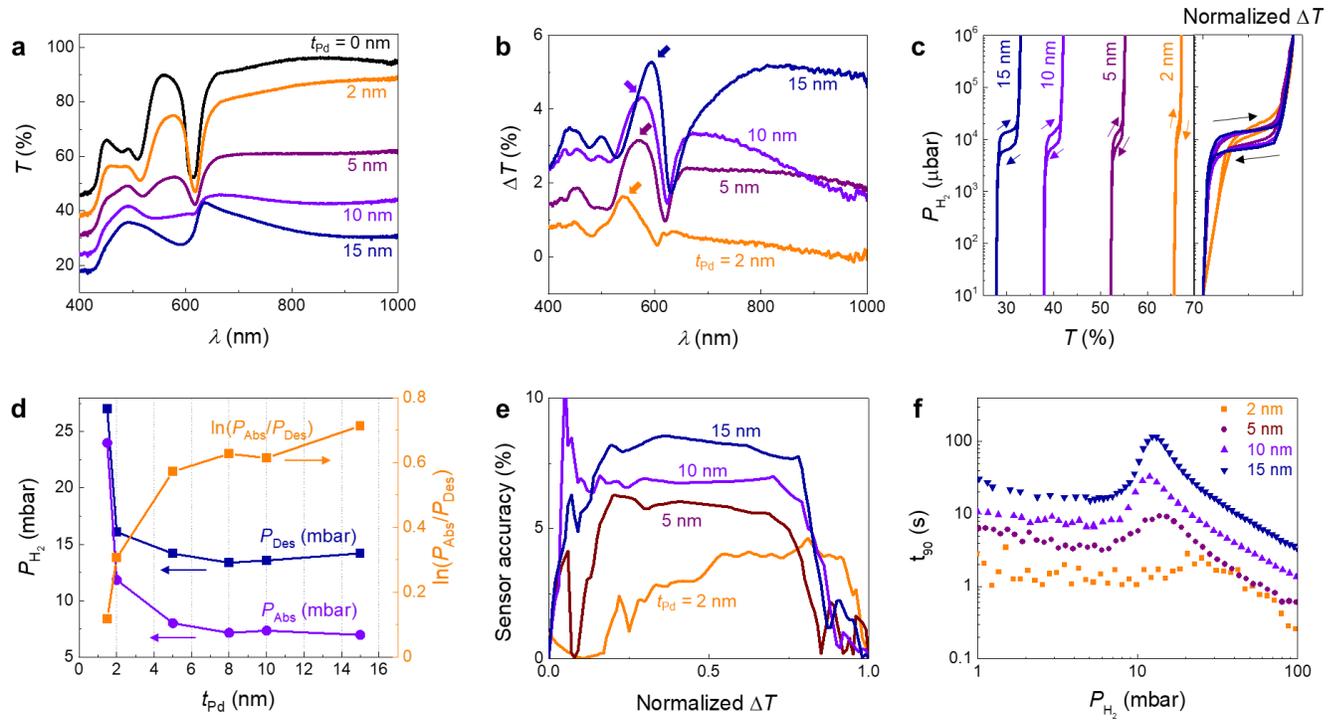

**Fig. 2 | Pd NP$^{50}_{t_{Pd}}$ sensors with different $t_{Pd}$.** (a) Experimental optical transmission spectra $T(\lambda)$ (at $P_{H_2}$ = 0 mbar) and (b) optical transmission changes $\Delta T(\lambda) = T_{1000\,mbar} - T_{0\,mbar}$ of NP with different $t_{Pd}$. Colored arrows indicate $\Delta T(\lambda)$ maxima. (c) Optical hydrogen sorption isotherm of NP with different $t_{Pd}$, extracted at $\Delta T(\lambda)$ maxima. The panel to the right shows corresponding normalized $\Delta T(\lambda)$ hydrogen sorption isotherm of NP with different $t_{Pd}$. Arrows denote the sorption direction. (d) Extracted plateau pressures for hydrogen absorption ($P_{Abs}$), desorption ($P_{Des}$), and $\ln(P_{Abs}/P_{Des})$ for different $t_{Pd}$. (e) Sensor accuracy at specific normalized $\Delta T$ readout over hydrogen pressure range of $10^1$ μbar to $10^6$ μbar. (f) Response time of NP sensors with pulse of hydrogen pressure from 1 mbar to 100 mbar.

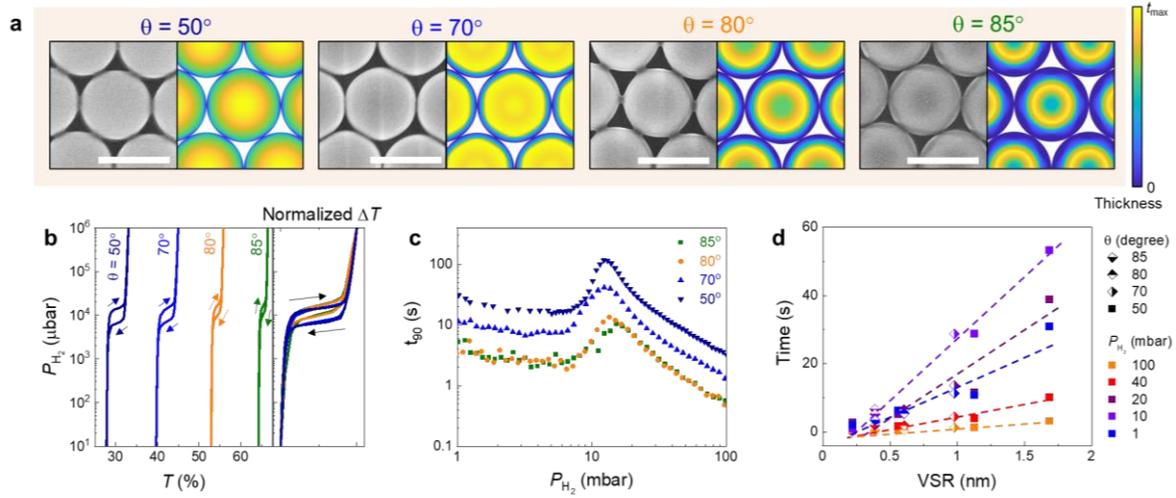

**Fig. 3 | Pd NP$_{15}^{\theta}$ sensors with different θ.** (a) Top-view SEM images (left) and corresponding simulated morphologies (right) of pure Pd NP samples, fabricated with vapor incident angle of θ = 50°, 70°, 80°, and 85°, respectively. Scale bars correspond to 500 nm, and the color bar shows the Pd thickness distribution in the simulated hemisphere cap. (b) Optical hydrogen sorption isotherm of NP with different θ, extracted at ΔT(λ) maxima. Arrows denote the sorption direction. The panel to the right shows corresponding normalized ΔT(λ) hydrogen sorption isotherm of NP with different θ. (c) Response time of NP sensors with pulse of hydrogen pressure from 1 mbar to 100 mbar. (d) Response time of NP samples as a function of volume-to-surface ratio of NP sensors, at different hydrogen pressure. Note that the optical isotherm data of NP$_{15}^{50}$ sample presented in Figs. 2c and 3b are identical, and response time data of NP$_{15}^{50}$ sample presented in Figs. 2f and 3c are identical.

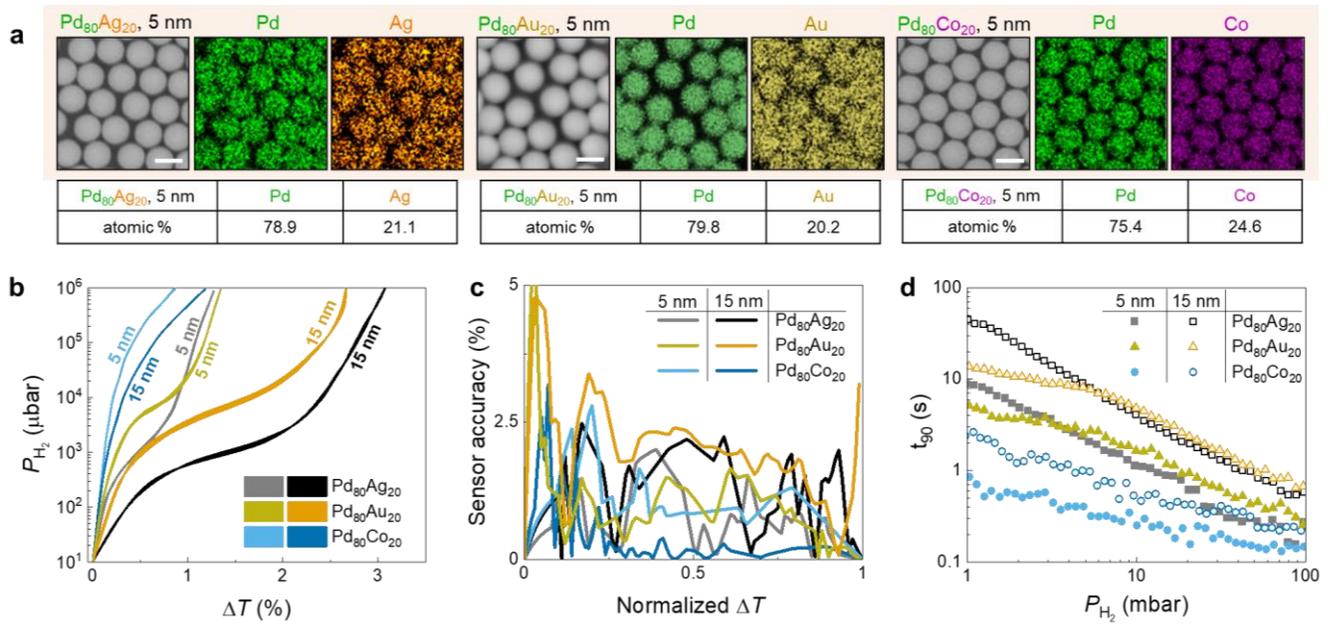

**Fig. 4 | Composite NP sensors.** (a) SEM and EDS images of $Pd_{80}Ag_{20}$, $Pd_{80}Au_{20}$, and $Pd_{80}Co_{20}$ ($t$ = 5 nm) composite NP samples. Scale bars correspond to 500 nm. Tables show the elemental atomic composition (at. %) of the $Pd_{80}Ag_{20}$ and $Pd_{80}Co_{20}$ NPs, which are consistent with the desired compositions. (b) Optical hydrogen sorption isotherm of composite NP, extracted at $\Delta T(\lambda)$ maxima. (c) Sensor accuracy of composite NP sensors at specific normalized $\Delta T$ readout over hydrogen pressure range of $10^1$ μbar to $10^6$ μbar. (d) Response time of NP sensors with pulse of hydrogen pressure from 1 mbar to 100 mbar.

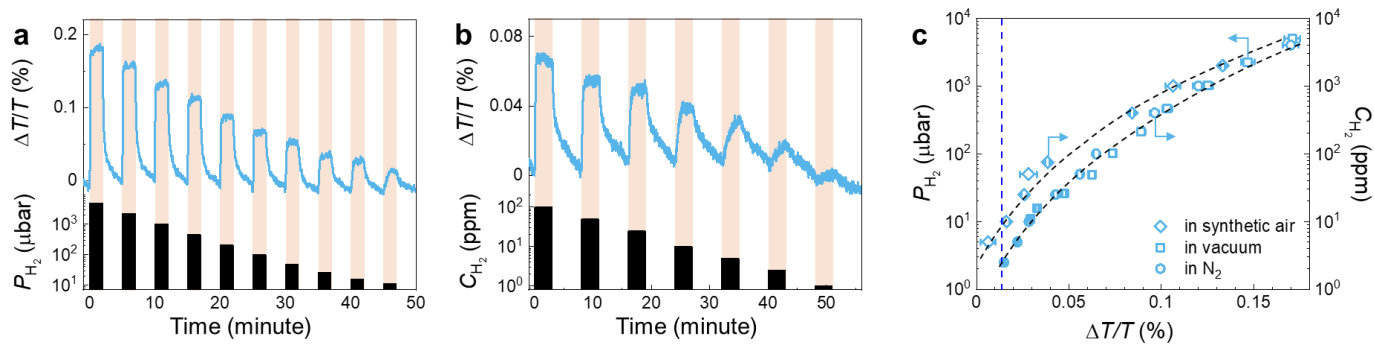

**Fig. 5 | Sensing performances of composite PdCo NP sensors.** (a) $\Delta T/T$ response of $Pd_{80}Co_{20}$ $NP_5^{50}$ sensors to stepwise decreasing hydrogen pressure in the 5000 – 11 μbar range, measured at 1.25 Hz sampling frequency in a vacuum chamber. (b) $\Delta T/T$ response of $Pd_{80}Co_{20}$ $NP_5^{50}$ sensors with different hydrogen concentrations ($C_{H_2}$) of 100 - 1 ppm, measured in flowing nitrogen (400 ml/min). Shaded areas denote the periods where the sensor is exposed to hydrogen. (c) Measured $\Delta T/T$ response as a function of hydrogen pressure/concentration derived from (a-b) and Supplementary Figs. 20–21. The blue dashed line denotes the defined LOD at $3\sigma \approx 0.013$ % (see Supplementary Section S9).

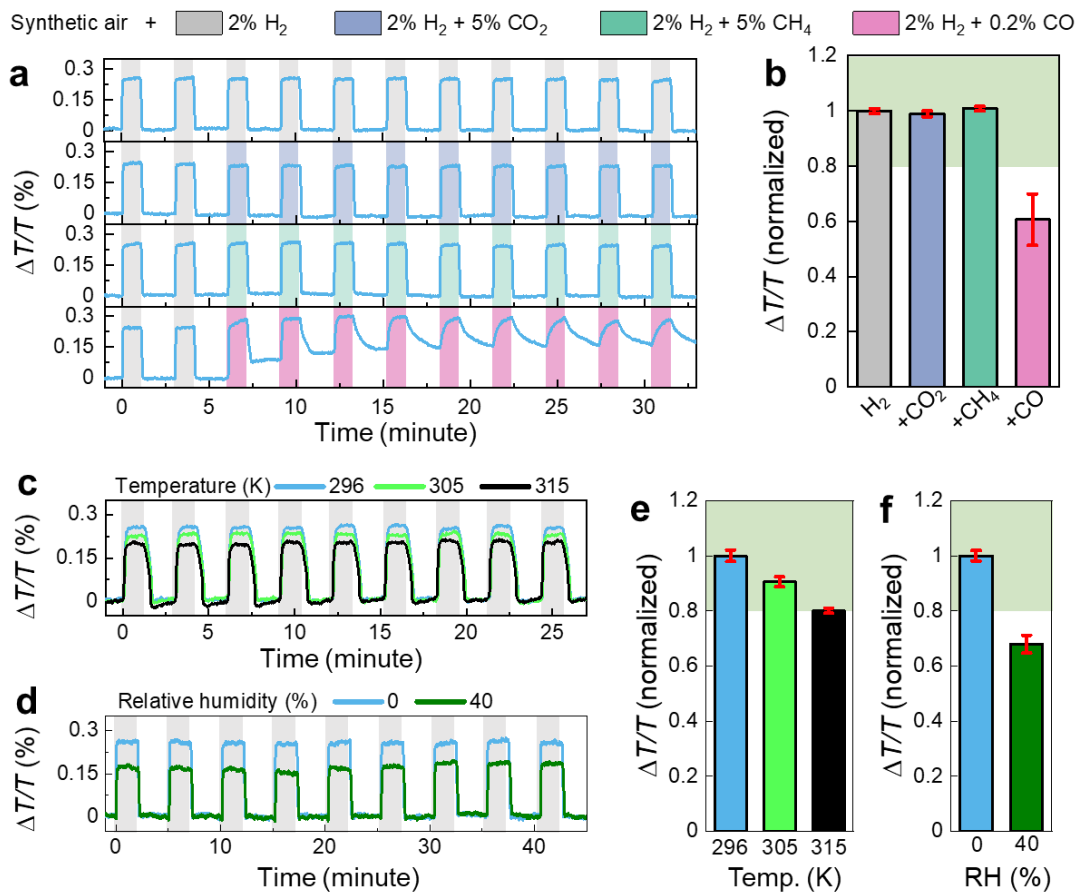

**Fig. 6 | Sensing performances of composite PdCo NP sensors.** (a) Time-resolved $\Delta T/T$ response of $Pd_{80}Co_{20}$ $NP_5^{50}$ to 2 pulses of 2% $H_2$ followed by 9 pulses of 2% $H_2$ + 5% $CH_4$, 2% $H_2$ + 5% $CO_2$, and 2% $H_2$ + 0.2% CO; and (b) normalized sensor signal to the one obtained with 2% $H_2$ in synthetic gas flow. The error bars denote the standard deviation from 9 cycles. (c) Time-resolved $\Delta T/T$ response of $Pd_{80}Co_{20}$ $NP_5^{50}$ to 9 pulses of 2% $H_2$ with different temperature and (e) normalized sensor signal one obtained with 2% $H_2$ in synthetic gas flow at 296 K. (d) Time-resolved $\Delta T/T$ response of $Pd_{80}Co_{20}$ $NP_5^{50}$ to 9 pulses of 2% $H_2$ with different relative humidity (RH) and (f) normalized sensor signal one obtained with 2% $H_2$ in dry condition. All measurements were performed using synthetic gas as carrier gas. The green shaded areas in (b), (d), and (e) indicate the ±20% deviation limit from the normalized $\Delta T/T$ response with 2% $H_2$.

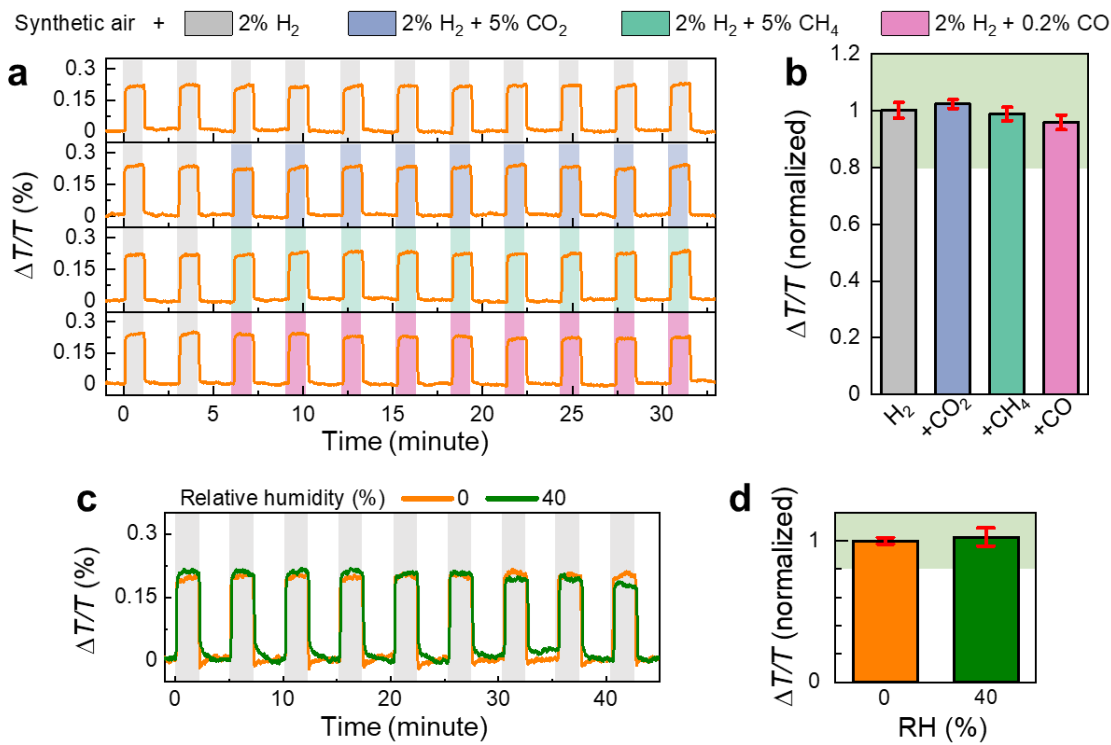

**Fig. 7 | Sensing performances of composite PdCo NP/PMMA sensors.** (a) Time-resolved $\Delta T/T$ response of $Pd_{80}Co_{20}$ $NP_5^{50}$/PMMA to 2 pulses of 2% $H_2$ followed by 9 pulses of 2% $H_2$ + 5% $CH_4$, 2% $H_2$ + 5% $CO_2$, and 2% $H_2$ + 0.2% CO; and (b) normalized sensor signal to the one obtained with 2% $H_2$ in synthetic gas flow. The error bars denote the standard deviation from 9 cycles. (c) Time-resolved $\Delta T/T$ response of $Pd_{80}Co_{20}$ $NP_5^{50}$/PMMA to 9 pulses of 2% $H_2$ with different relative humidity (RH) and (d) normalized sensor signal one obtained with 2% $H_2$ in dry condition. All measurements were performed using synthetic gas as carrier gas. The green shaded areas in (b) and (d) indicate the ±20% deviation limit from the normalized $\Delta T/T$ response with 2% $H_2$.

# Sub-second and ppm-level Optical Sensing of Hydrogen Using Templated Control of Nano-hydride Geometry and Composition

Hoang Mai Luong*[1], Minh Thien Pham[1], Tyler Guin[2], Richa Pokharel Madhogaria[3], Manh-Huong Phan[3], George Keefe Larsen*[2], and Tho Duc Nguyen*[1]

[1]*Department of Physics and Astronomy, University of Georgia, Athens, Georgia 30602, USA.*

[2]*National Security Directorate, Savannah River National Laboratory, Aiken, South Carolina 29808, USA.*

[3]*Department of Physics, University of South Florida, Tampa, Florida 33620, USA.*

*E-mail: hoanglm@uga.edu, george.larsen@srnl.doe.gov, ngtho@uga.edu

# Table of Contents

# S1. The thickness distribution simulation of nano-patchy (NP) samples

## S1.1. Methods

We estimate the thickness distribution, surface, and volume of the hemisphere cap, based on a simple simulation including a uniform vapor flux approaches in direction $\hat{l}(\theta_0, \varphi_0)$ to an array of hexagonal close-packed nanosphere with diameter $D = 500$ nm (Figure S1a).[1,2] The thickness distribution of the nanosphere $O$ (highlighted in orange) is calculated by considering the shadowing effects from 36 nearest-neighbor nanospheres. The surface of nanosphere $O$ is broken down into smaller surface elements, and each of them is labelled by the polar coordinate of its center $(\theta, \varphi)$, where $\theta = i\Delta\theta$, $\varphi = j\Delta\varphi$, $\Delta\theta = \Delta\varphi = 0.5°$, $i = 0, 1, \ldots, 360, j = 0, 1, \ldots, 720$.

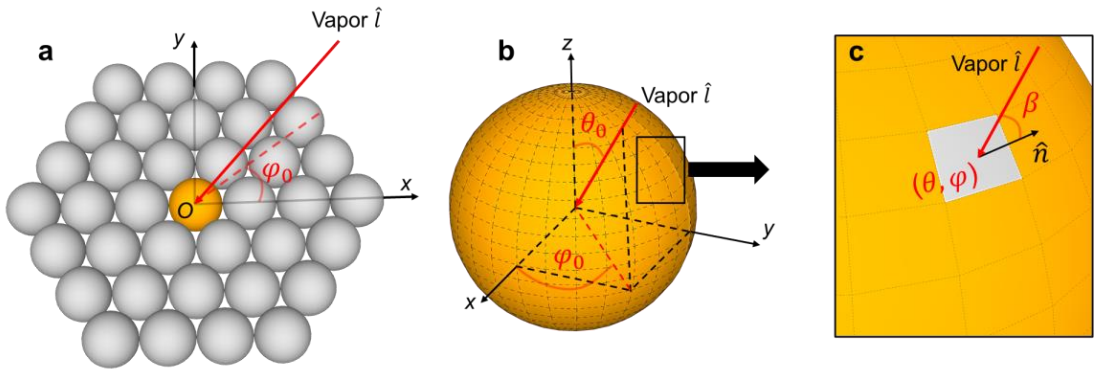

**Figure S1.** (a) A cartoon illustrates the vapor deposition on an array of nanospheres. (b) The surface of nanosphere $O$ split into several surface elements. (c) A surface element on the nanosphere $O$.

Our simulation is based on these following assumptions:

- Only the shadowing effect and material accumulation are considered. Other physical processes, such as surface diffusion or material penetration, are neglected.
- The deposition at different surface elements happens simultaneously as long as they are directly exposed to vapor.
- The as-deposited film is non-porous and is uniform within each surface element.

In our experimental metal deposition, the substrate holder was rotated azimuthally at a constant rate of 30 rpm to thoroughly cover the top surface of nanosphere. Therefore, in order to mimic this process, we break down the simulation into 3600 steps. We start at $\varphi_0 = 0°$ (step index $k_i = 1$), a 0.1° azimuthal rotation of $\hat{l}(\theta_0, \varphi_0)$ happens at the end of each step, until a round of azimuthal rotation is completed ($\varphi_0 = 359.9°$) (step index $k_f = 3600$). In each simulation step, the thickness at each surface element $h(\theta, \varphi)$ is updated,

$$h_{k+1}(\theta, \varphi) = h_k(\theta, \varphi) + \Delta h_k. \tag{S1}$$

The change of thickness for each step $\Delta h_k$ is determined by whether surface element $(\theta, \varphi)$ is directly exposed to the vapor flux or not:

- If the surface element $(\theta, \varphi)$ is under the shadow of other structures (deposited materials on neighboring bead in the previous steps are also considered), then $\Delta h_k = 0$.

- If the surface element $(\theta, \varphi)$ can receive vapor, then $\Delta h_k = \frac{\Delta m}{\rho \cdot S(\theta, \varphi)}$, where $\Delta m$ is the mass of material deposited on surface element $(\theta, \varphi)$ within time $\Delta t$, $\rho$ is the density of the material, and $S(\theta, \varphi)$ is the area of surface element $(\theta, \varphi)$.

The vapor flux $\Phi$ is defined as,

$$\Phi = \frac{\Delta m}{\Delta t \cdot S_N} = const, \tag{S2}$$

where $S_N$ is the projection of $S(\theta, \varphi)$ onto the plane perpendicular to the vapor flux $\hat{l}(\theta_0, \varphi_0)$. With $\beta$ is the angle between $\hat{l}$ and the surface normal vector $\hat{n}$ (Figure R1c), $S_N$ can be written as,

$$S_N = S(\theta, \varphi) \cos\beta. \tag{S3}$$

Combining (S1)-(S3), we achieve,

$$\Delta h_k = \frac{\Phi}{\rho}\Delta t \cos\beta. \tag{S4}$$

For the simulation of $\text{NP}_{t_\text{Pd}}^{\theta_0}$ sample, in each step k, we set $\frac{\Phi}{\rho}\Delta t = \frac{t_\text{Pd}}{k_f} = \frac{t_\text{Pd}}{3600}$ (nm). After the last simulation step of $k_f$, a 3-dimension hemisphere cap are rendered based on the thickness distribution on the nanosphere $O$, and the volume and surface area of the patchy particle are simply obtained by using double and triple integral built-in function of MATLAB. The results are presented in Figures S2 and S3.

## S1.2. The calculation of volume-to-surface ratio of nano-patchy (NP) samples with different vapor incident angles (θ)

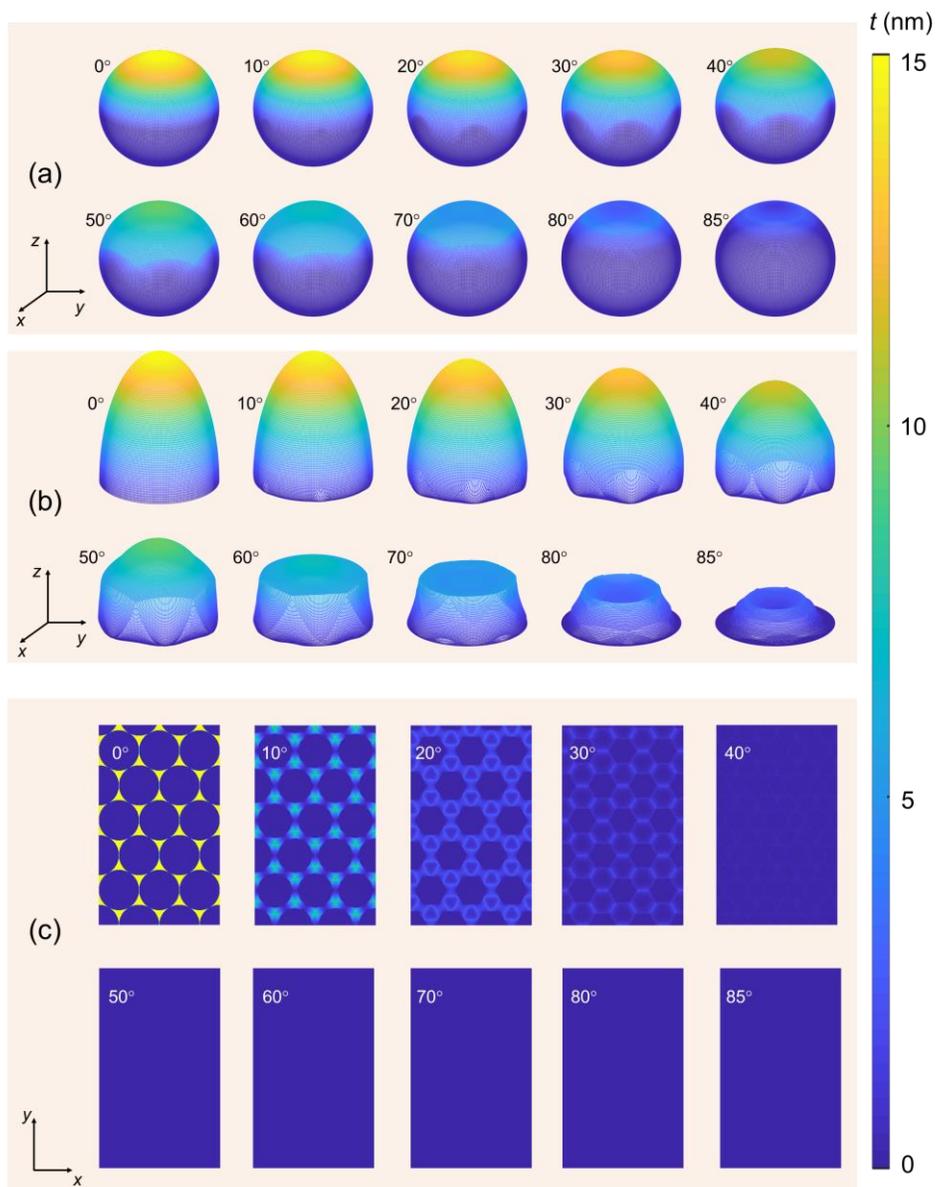

**Figure S2.** (a) Simulated NP morphologies with (b) corresponding thickness distribution on polystyrene (PS) nanosphere projected to a flat surface, and (c) corresponding nano-patterns formed on glass substrate, with different vapor incident angles, θ (the deposited thickness is fixed at $t = 15$ nm). Note that in (b) z-axis is not-to-scale with x-axis and y-axis.

Figure S2 presents the simulated morphology of NP under different vapor incident angles, θ (at *t* = 15 nm), which shows a decrease maximum deposited thickness on PS nanosphere when θ increases. We further estimate the surface area and the volume of NP with respect to the vapor incident angle as shown in Figure S3a. While the volume of deposited material gradually decreases when θ increases, we observe a sharp drop of surface area when θ ≥ 70°. Volume-to-surface ratio (VSR, orange curve in Figure S3b) is consequently calculated using these values in Figure S3a. We also estimate VSR of the NP in the case it is projected into a flat surface, which show about two times large of VSR in comparison with the one on PS nanosphere (pink curve in Figure S3b), regardless of θ.

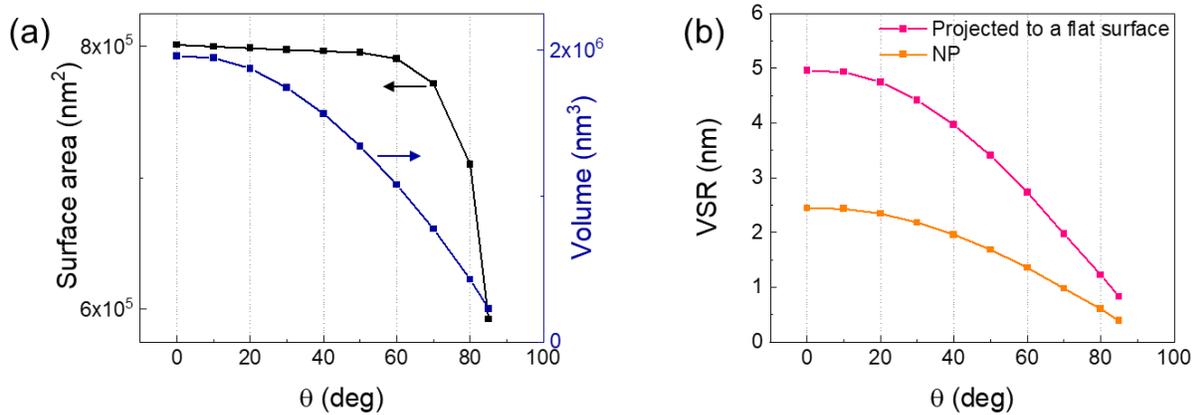

**Figure S3.** (a) Estimated surface area and volume of NP sample with respect to the vapor incident angle θ and (b) estimated VSR of thickness distributions in Figure S2a and S2b.

## S1.3. The calculation of volume-to-surface ratio of nano-patchy (NP) samples with different deposited thicknesses (θ = 50° is fixed)

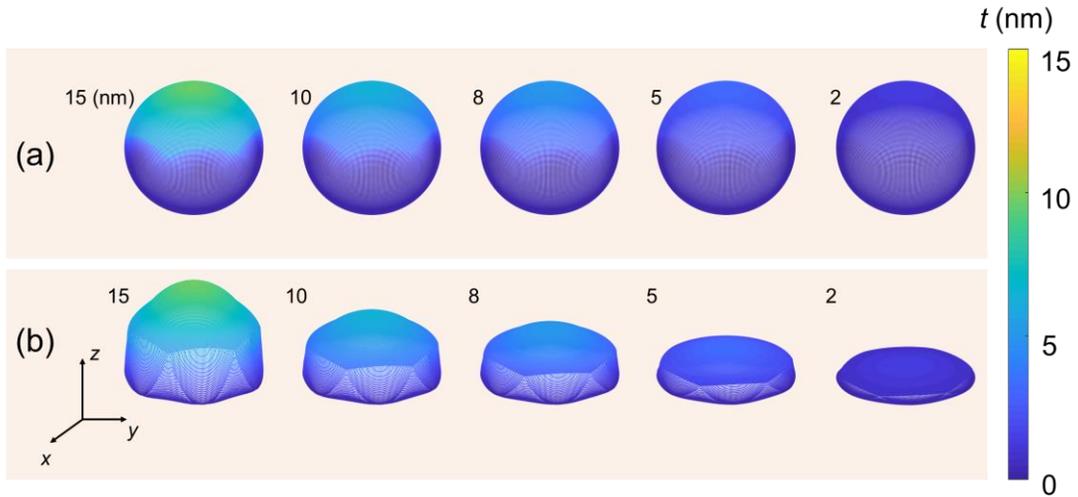

**Figure S4.** (a) Simulated NP morphologies with different deposited thickness $t$ (the vapor incident angle is fixed at θ = 50°) and (b) corresponding deposited thickness distribution on polystyrene (PS) nanosphere projected to a flat surface. Note that in (b) z-axis is not-to-scale with $x$-axis and $y$-axis.

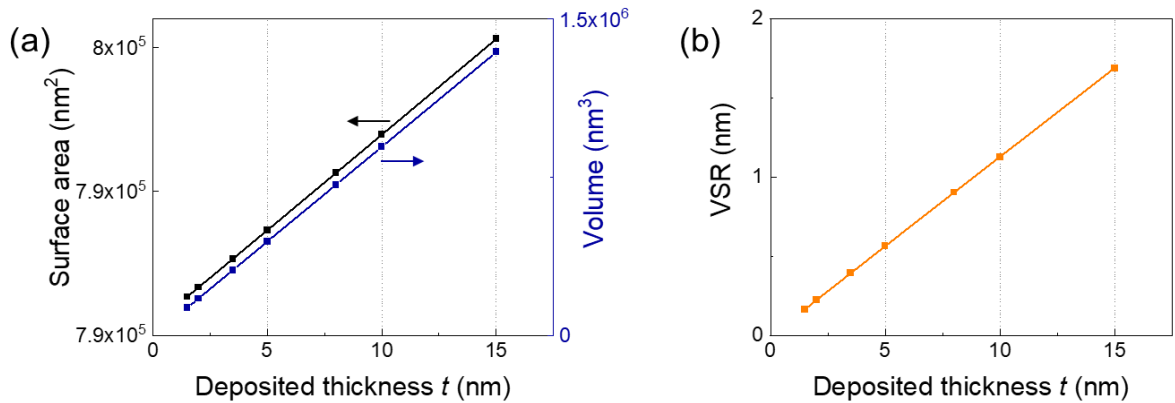

**Figure S5.** (a) Estimated surface area and volume of NP sample with respect to the deposited thickness $t$ and (b) estimated VSR of thickness distributions in Figure S4a.

## S2. Additional structural characterization

### S2.1. Scanning electron microscopy (SEM) micrographs of NP samples

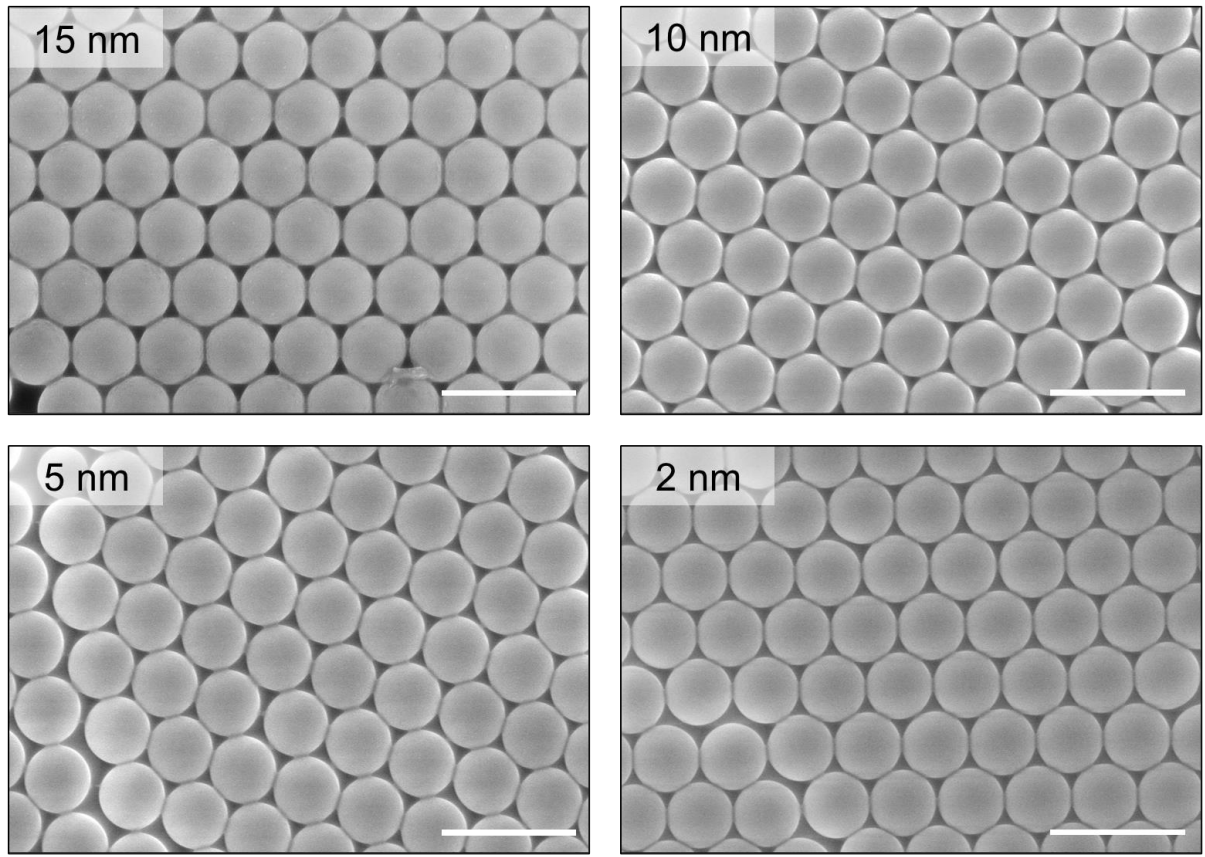

**Figure S6.** Top-view scanning electron microscope (SEM) images of Pd $\text{NP}_t^{50}$ sample with different deposited thicknesses ($t$ = 2, 5, 10 and 15 nm). Scale bars, 1000 nm.

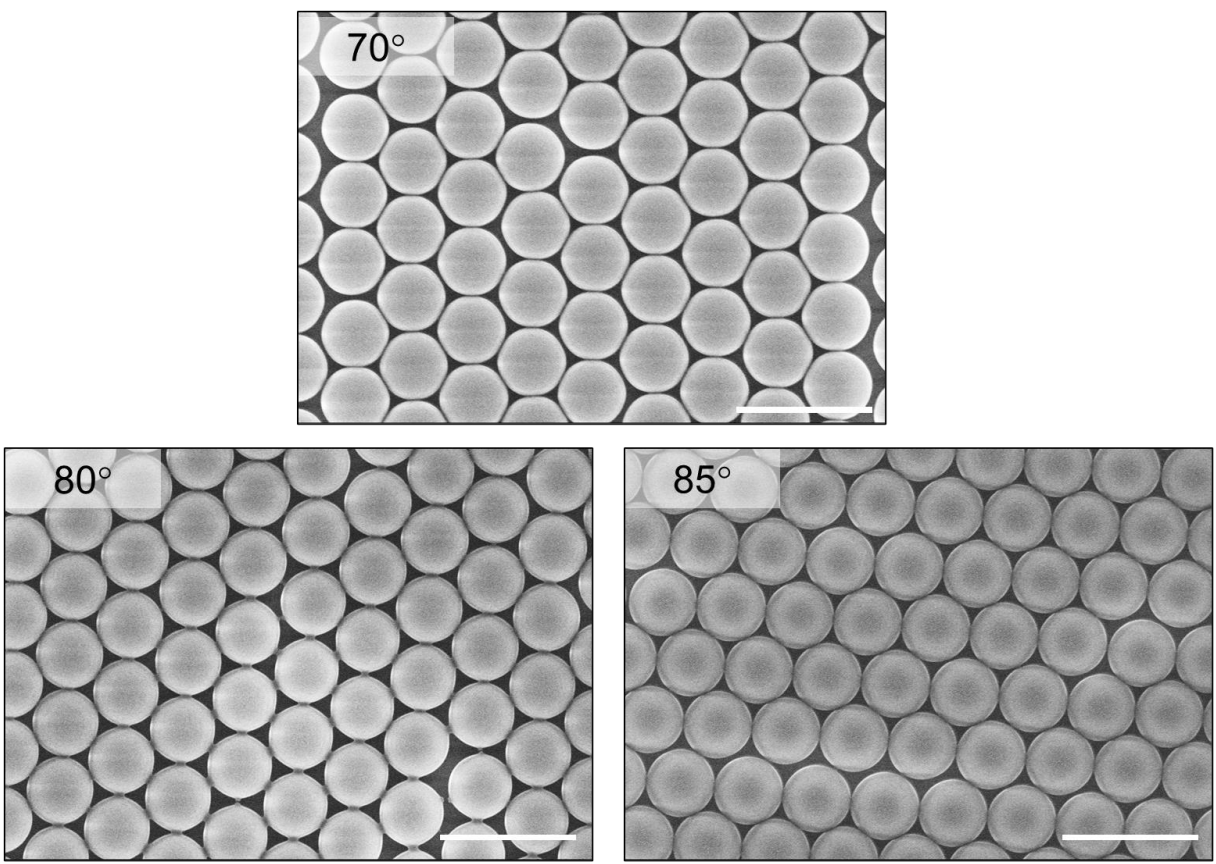

**Figure S7.** Top-view scanning electron microscope (SEM) images of Pd NP$_{15}^{\theta}$ samples with different θ = 70°, 80°, and 85°. Scale bars, 1000 nm.

## S2.2. Morphological transition of Pd, PdAg, PdAu, and PdCo NPs with different deposited thickness

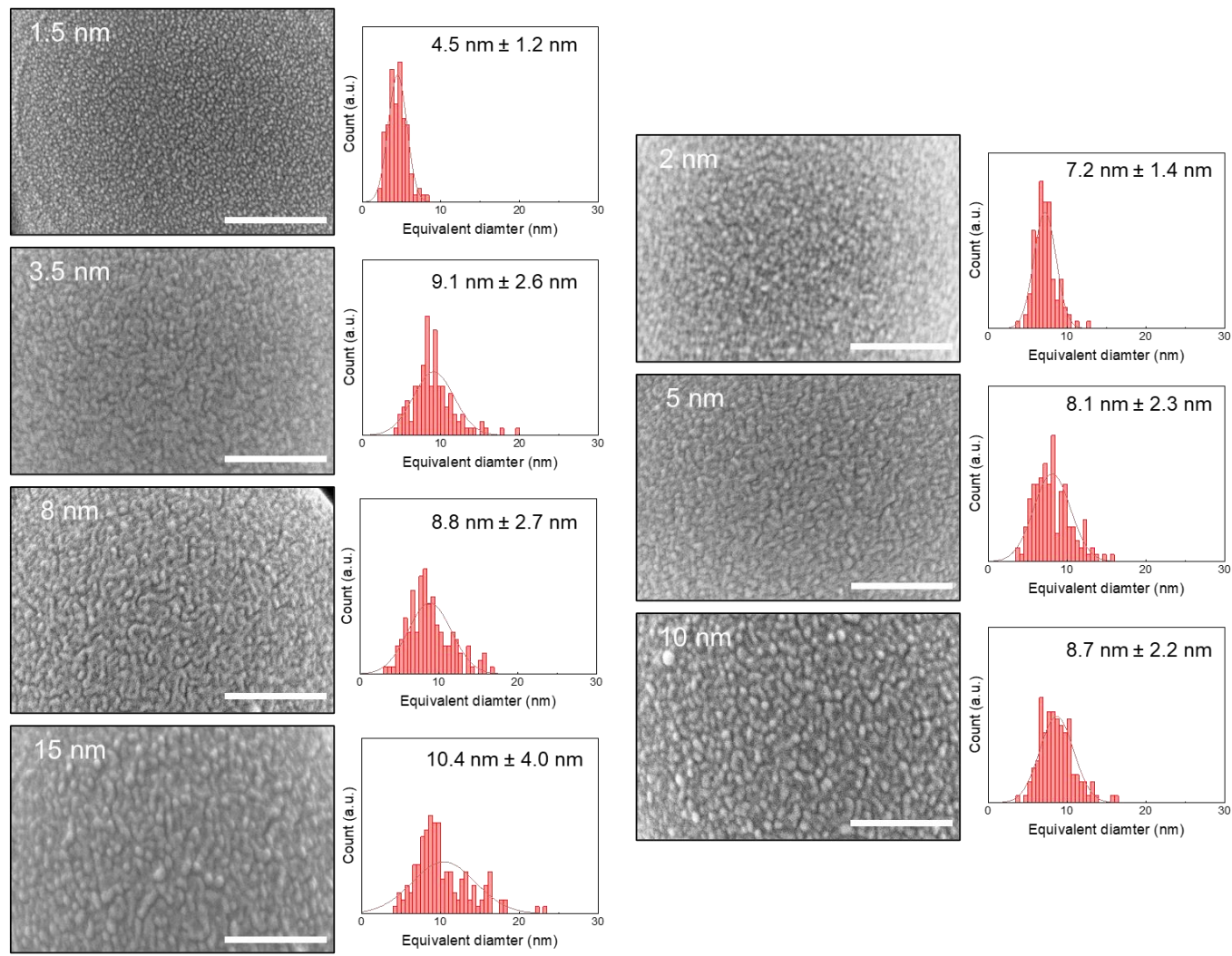

**Figure S8.** Ultra-high-resolution SEM images and associated grain size analysis, showing morphology differences of Pd NP with different deposited thicknesses ($t_{Pd}$ ranges from 1.5 to 15 nm). Scale bar: 100 nm.

The morphological transition of Pd $NP_{t_{Pd}}^{\theta}$ with increasing $t_{Pd}$ was observed using ultra-high-resolution SEM (SU-9000, Hitachi), as revealed in Figure S8. The morphology of $NP_{1.5}^{50}$ contains many sub-10-nm granules, and these granules cover fully the top surface of the polystyrene nanosphere. The size of the granules grows when $t_{Pd}$ = 2 nm, and the coalescence via a neck (or bridge) connection between neighboring clusters can be noticed at the thickness of $t_{Pd}$ = 3.5 nm. A continuous film is formed at a thickness of 5 nm, as these bridge connections are successively grown (a cross-sectional SEM image and EDS elemental maps of $NP_5^{50}$ can be found in Figure 1d). Once the continuous film is shaped at a thickness of 5-nm, another distinct coalescent mechanism is observed: the size of the clusters increases as the thickness increases, associated with the reduction of the cluster density.

In order to quantify these effects, we define a volume fraction, $V_f$, such that

$$V_f = V_{dep}/V_{meas} \tag{S5}$$

$$V_{dep} = \frac{4\pi}{6}\left((R + t_{dep})^3 - R^3\right) \tag{S6}$$

$$V_{meas} = \frac{4\pi}{6}((R + t_{meas})^3 - R^3), \tag{S7}$$

where $V_{dep}$ is the hemispherical volume of material deposited based on a deposition of material thickness, $t_{dep}$, (as determined by the quartz crystal microbalance during the fabrication) onto a sphere of radius, $R$ ($R$ = 500 nm in this case). $V_{meas}$ would be the volume of a hemispherical shell on a sphere of radius $R$ if the thickness of the shell, $t_{meas}$, matched the average grain size measured in the high-resolution SEM images (Figure S8 above). Figure S9a shows the calculated $V_f$ for the different Pd films, and while the standard deviation, σ, is rather large for the calculations, a general trend can be observed. Consistent with the qualitative description above, films with $t_{Pd} \geq 3.5$ nm have a volume fraction that exceeds the percolation threshold $p_c$ for bond percolation ($p_c$ = 0.347296), and films with $t_{Pd} \geq 5$ nm have a volume fraction that exceeds $p_c$ for site percolation ($p_c$ = 0.5).[3] In this case, we consider a film to be island-like if $V_f < 0.5$ and film-like if $V_f \geq 0.5$.

As shown in Figure S10, different alloying elements (Ag, Au, Co) impact the nanoparticle morphology differently. Therefore, it is also interesting to compare the effects of the different alloying elements on $V_f$. Figure S9b presents the $V_f$ for the different Pd alloy films. The general trend for $V_f$ is PdAg < Pd < PdAu <

PdCo, with Pd ≈ PdAg and PdAu ≈ PdCo more loosely. It is worth noting that these films all have $V_f \geq 0.5$, but also have large σ values, where Pd and PdAg have the largest variations.

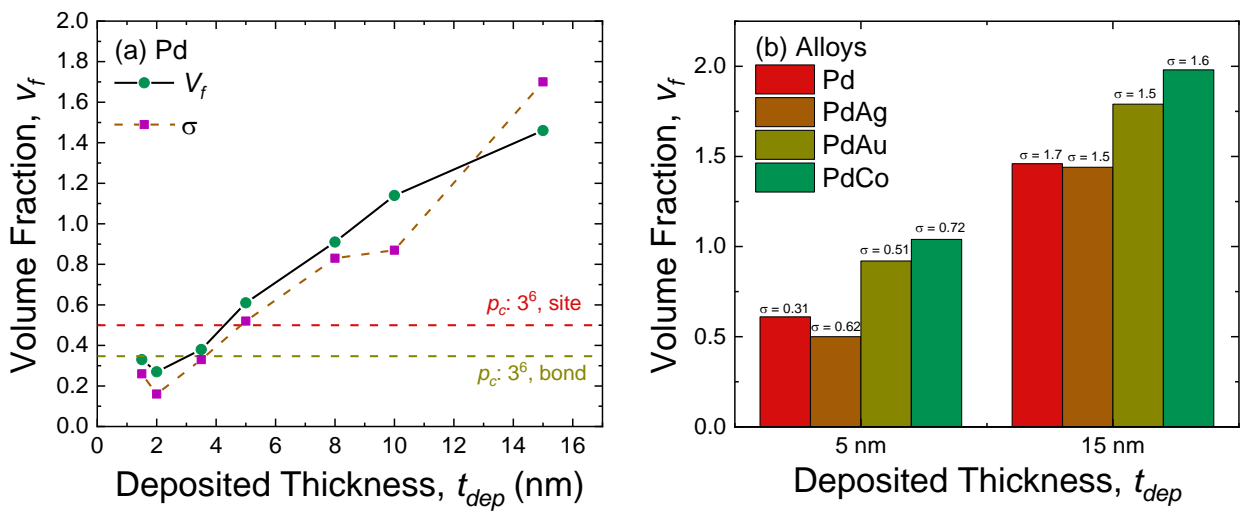

**Figure S9.** Calculated volume fraction, $V_f$, of (a) the Pd samples versus deposited thickness and (b) the different Pd alloys for $t_{dep}$ = 5 and 15 nm. The values of standard deviation, σ, for the $V_f$ calculations are illustrated by the pink squares and placed above the bars in (a) and (b), respectively.

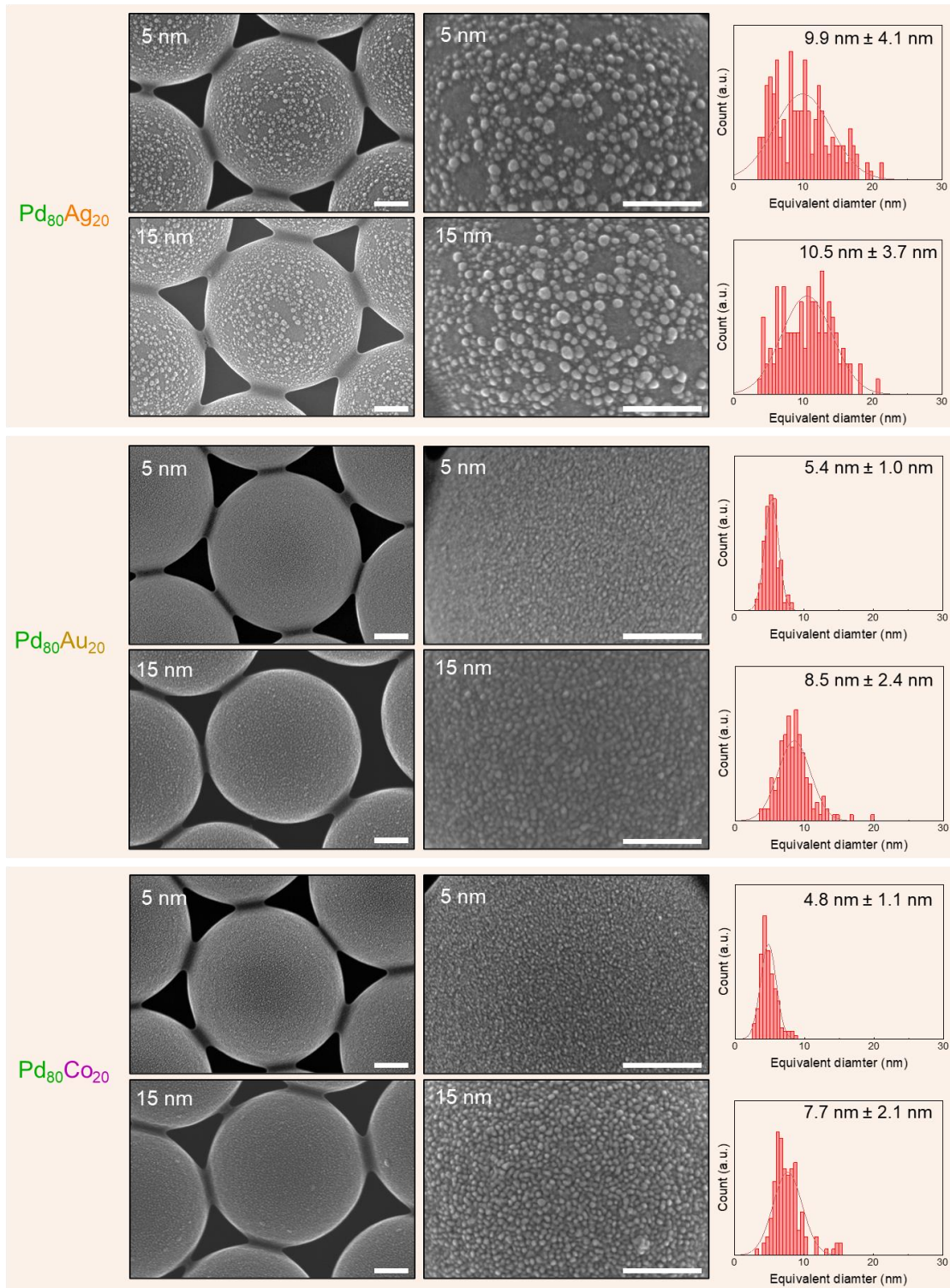

**Figure S10.** Ultra-high-resolution SEM images and associated grain size analysis, showing morphology differences of NP with different deposited thicknesses (5 and 15 nm) and compositions. Scale bar: 100 nm.

## S2.3. Size dependence of $\ln(P_{Abs}/P_{Des})$, $P_{Abs}$, and $P_{Des}$ in $NP_{t_{Pd}}^{50}$ films

As shown in Figure 2d in the main text, $\ln(P_{Abs}/P_{Des})$, $P_{Abs}$, and $P_{Des}$ display a size dependent behavior, where a noticeable transition occurs for $t_{Pd} < 5$ nm. This transition in behavior matches the transition between island-like to film-like morphologies described in Subsection S2.2 above. Further, the size-dependent effects can be described by a size-dependent critical temperature, $T_c$, and an increasing effect of subsurface sites for smaller particles.

The size dependence of hysteresis in nanoparticles, which is quantitively expressed as $\ln(P_{Abs}/P_{Des})$, has been comprehensively analyzed by Griessen et al. and a robust scaling law has been proposed.[4] Based on a simple lattice gas model, the hysteresis was found to only depend on the ratio of $T/T_c$, where $T_c$ is the critical temperature at which the hysteresis vanishes. The full-spinodal line for hysteresis is given by:

$$\ln\left(\frac{P_{us}}{P_{ls}}\right) = 8\frac{T_c}{T}z + 4\ln\left(\frac{1-z}{1+z}\right), \tag{S8}$$

where $z = \sqrt{1-(T/T_c)}$ and $P_{us}$ and $P_{ls}$ are upper and lower spinodal pressures, respectively. Griessen et al. found that the hysteresis behavior of a wide variety of Pd nanoparticles fell between the full spinodal hysteresis line given by Equation S8 and 45% of the full spinodal hysteresis value. The size dependence of $T_c$ was found to be well represented by:

$$T_c = A - (B/L), \tag{S9}$$

where $A$ and $B$ are fitting parameters and $L$ is the size of a nanocube in nanometers. Using $T = 300$ K and Equation S8 we can similarly extract $T_c$ values as a function of $t_{Pd}$ for the $NP_{t_{Pd}}^{50}$ films and determine the scaling relationship with $t_{Pd}$. The results assuming a full spinodal hysteresis are presented in Figure S11. $T_c$ varies from 324 - 382 K with increasing size and is well fit by Equation

S9 with $A = 387.054$ and $B = 90.1308$. While not presented here, the results for 45% spinodal hysteresis (0.45 is added as a multiplicative factor to the right-hand side (RHS) of Equation S9) show an increase $T_c$ to 341 to 445 K, and are fit by Equation S9 with $A = 453.02$ and $B = 159.995$. These values agree well with the results of Griessen et al. and provide further evidence of the universality of the relationship derived therein, as the fitting here encompasses the transition between island-like and film-like morphology. Thus, the shrinking hysteresis observed in the $NP_{t_{Pd}}^{50}$ films is due to a decreasing $T_c$ value with decreasing $t_{Pd}$.

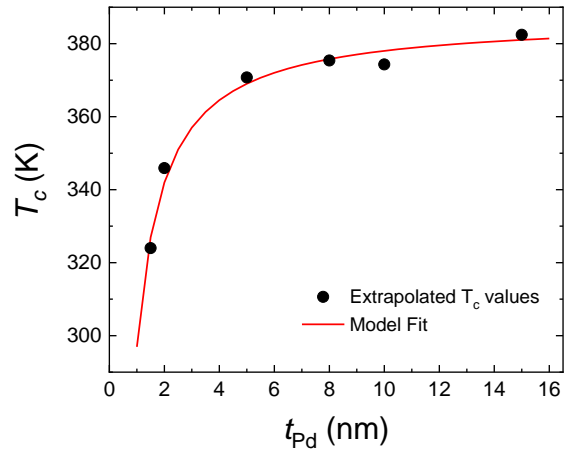

**Figure S11.** $T_c$ values as a function of Pd thickness. The data points were obtained from experimental values using Equation S8. The model curve was generated using Equation S9 with fitting parameters.

The $\ln(P_{Abs}/P_{Des})$ values can also be used to understand the effects of subsurface sites on $P_{Abs}$, and $P_{Des}$. According to Sachs et al. the fraction $c_t$ of subsurface sites can be determined given the well-supported assumption that hydrogen in subsurface sites does not transform to a hydride phase:

$$c_t = 1 - \frac{c_\beta^{nano} - c_\alpha^{nano}}{c_\beta^{bulk} - c_\alpha^{bulk}}, \tag{S10}$$

where the superscripts denote the nanoparticle or bulk concentrations and the subscripts denote the concentrations at the boundaries of the $\alpha$ and $\beta$ phases.[5] As described in the main text, the hysteresis can be explained by the thermodynamics of an open, coherent two-phase system. According to Schwarz and Khachaturyan, the absorption and desorption plateau pressures may then be calculated by:

$$\ln\left(\frac{P_{Abs}}{P_{Des}}\right) = \left(\frac{4\Omega\, G_s \frac{1+v}{1-v}\varepsilon_0^2(c_\beta - c_\alpha)}{kT}\right), \quad (S11)$$

where the volume of one hydrogen atom in Pd, $\Omega = 2.607$ Å$^3$, the shear modulus of Pd, $G_s = 47.7 \times 10^9$ Pa, the Poisson number $v = 0.385$, the change in lattice constant, $\varepsilon_0 = 0.063$, and Boltzmann constant $k$.[6] Using $c_\alpha^{bulk} = 0.008$ and $c_\beta^{bulk} = 0.0607$, Equation S11 gives $\ln(P_{Abs}/P_{Des}) = 0.62$, which is in excellent agreement with the experimentally measured values for $t_{Pd} \geq 5$ nm as described in the main text. By combining Equations S10 and S11 and using the experimentally determined $\ln(P_{Abs}/P_{Des})$ values, the fraction of subsurface sites, $c_t$, can be calculated for each $t_{Pd}$ value. The values are $< 0.10$ for $t_{Pd} \geq 5$ nm but rapidly increase for $t_{Pd} < 5$ nm (Figure S12). The volumes of the bulk and subsurface regions in the film-like hemispherical patchy caps ($t_{Pd} \geq 5$ nm) may be respectively estimated as:

$$V_b = \frac{4\pi}{6}((R + t_{Pd} - h)^3 - R^3), \quad (S12)$$

$$V_{ss} = \frac{4\pi}{6}((R + t_{Pd})^3 - (R + t_{Pd} - h)^3), \quad (S13)$$

$$V_{tot} = V_{ss} + V_b, \quad (S14)$$

where $V_{tot}$ is the total volume, the radius of the PS bead template is $R = 500$ nm, and $h$ is the thickness of the subsurface layer. Values for $h$ in the literature range from 0.3 - 4 nm, depending on the structure and Pd crystalline surface.[4] By setting $c_t = V_{ss}/V_{tot}$ and solving for $h$, we find that $h = 0.4$ nm for $t_{Pd} \geq 5$ nm, which agrees very well with the literature values.

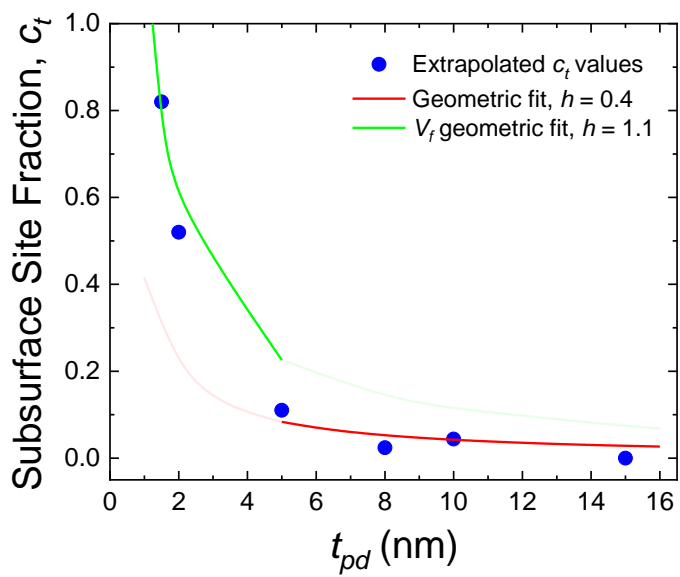

**Figure S12.** Subsurface sites as a function of Pd thickness. The data points were obtained from experimental values using Equation S11. The model curves were generated using Equations S12 – S14 with $h = 0.4$ nm or by volume fraction modification with $h = 1.1$ nm.

As described in Subsection S2.2 above, the Pd films transition from a film-like morphology to island-like morphology for $t_{Pd} < 5$ nm. In order to estimate the subsurface site layer thickness, we modify our assumptions such that $c_t = (1/V_f)(V_{ss}/V_{tot})$ and by using $t_{meas}$, instead of $t_{Pd}$ in Equations S12 – S14, where $V_f$ and $t_{meas}$ are defined in Subsection S2.2. In this case, we find h = 1.1 nm provides a good fit for $t_{Pd} < 5$ nm. It is worth noting a spherical calculation of $h$ based on $t_{meas}$ indicates $h = 2$ nm and 1.6 nm for $t_{Pd} = 1.5$ nm and 2 nm respectively.

Evaluating the experimental data through Equations S10 – S13 enabled quantification of the increasing relative volume of subsurface sites as the Pd film thickness decreases in the patchy particles. The importance of these sites on the thermodynamics of adsorption increases as the particle size gets smaller. Wadell et al. suggested that the behavior of the enthalpy of hydride formation as Pd nanoparticles shrink is a combination of the competitive effects of surface tension and subsurface sites.[7] That is, the surface tension contribution decreases as particles shrink, while the relative number subsurface sites, which have a higher absorption energy than bulk sites, increases. To quantify these effects, they developed an analytical Langmuirian-type model, where the chemical potentials of H atoms in the bulk hydride and subsurface volumes respectively are:

$$\mu_b^H = -E_0 + kT \ln\left(\frac{\theta_b}{1-\theta_b}\right) + \frac{2\gamma\Omega}{r}, \tag{S15}$$

$$\mu_{ss}^H = -E_0 - \Delta E + kT \ln\left(\frac{\theta_{ss}}{1-\theta_{ss}}\right) + \frac{2\gamma\Omega}{r}. \tag{S16}$$

$\theta_b$ and $\theta_{ss}$ are the H coverages of the bulk and subsurface sites, $E_0$ is the gain in energy during the hydride formation at R $\to \infty$, $\Delta E > 0$ is the energy difference between the bulk and subsurface sites, $\gamma$ is the surface tension (0.2 eV/Å$^2$ for Pd), and $\Omega$ is the partial volume of hydrogen in the hydride phase ($\Omega$ = 2.607 Å$^3$). The last terms on the RHS of Equations S15 and S16 are the surface tension term, and for the Pd patchy system $r = R + t_{Pd}$, with $R$ = 500 nm. At equilibrium, the chemical potential of H$_2$ molecules in the gas phase is:

$$\mu_{H_2} = kT \ln\left[\left(\frac{P}{(kT^{5/2})}\right)\left(\frac{2\pi\hbar^2}{m}\right)^{3/2}\right], \tag{S17}$$

$$\mu_{H_2} = 2\mu_b^H = 2\mu_{ss}^H, \tag{S18}$$

where $m$ is the mass of a H$_2$ molecule and $\hbar$ is the reduced Planck's constant. Note that this expression takes only the translation partition function into account, and the minor rotational and vibrational contributions are ignored. The average coverage of the bulk and subsurface sites is given by:

$$\langle\theta\rangle = \frac{V_b}{V_{tot}}\theta_b + \frac{V_{ss}}{V_{tot}}\theta_{ss}. \tag{S19}$$

By setting $\langle\theta\rangle = 0.5$, using the value for $h$ determined above for ($h = 0.4$ nm), selecting $\Delta E = 2.6$ kJ/mol H, and using the experimentally measured absorption plateau pressures at $T = 300$ K, Equations S15 – S19 can be used to solve for $\theta_b$, $\theta_{ss}$, and $E_0$ for the $t_{Pd} > 5$ nm values. The fittings result in average values of $\theta_b = 0.48 \pm 0.01$, $\theta_{ss} = 0.971 \pm 0.001$, and $E_0 = 19.6 \pm 0.1$ kJ/mol H. The obtained value of $E_0 \approx 20$ kJ/mol H in excellent agreement with literature values.[4,7] On the other hand, $\theta_{ss}$ is higher than expected ($\theta_{ss} \times 0.607 = 0.44$ H/Pd, typically), and thus could indicate higher H capacity in surface shell sites facilitated by the polymer substrate. Similar effects were seen in Pd nanocubes covered with a metal organic framework.[8] It is interesting to note that the subsurface site effect is more pronounced in the Pd patchy particles than it is in many other Pd nanoparticles. This is because the surface curvature changes much more slowly than the volume to surface ratio in the Pd patches. Hence, Wadell et al. observed a relatively constant enthalpy of hydride formation for decreasing particle size in their spherical nanoparticles.

In addition, we note that the hydrogen absorption and desorption isotherm pressures above the critical thickness ($t_{Pd} \approx 5$ nm) are $P_{Abs} \approx 13$-$14$ mbar and $P_{Des} \approx 7$ mbar, respectively, both of which are appreciably smaller than those of Pd thin films and Pd nano-particles on flat substrates ($P_{Abs} \approx 22$-$25$ mbar and $P_{Des} \approx 10$-$11$ mbar).[9,10] The lowering of plateau pressures may arise from in-plane strain imposed by the curvature of the substrate, which increases the long-range attractive H–H interactions.[11,12] Alternatively, a reduced adhesive force between the Pd structures and the softer PS substrate may also contribute to the lowering of the plateau pressures relative to Pd structures supported on rigid substrates.[12]

## S2.4. Power law fitting of response time in the $\text{NP}^{50}_{t_{\text{Pd}}}$ films

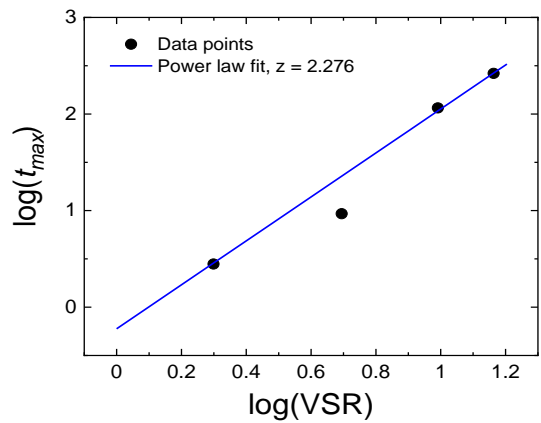

**Figure S13**. Power law fitting ($t_{\max} = a \times \text{VSR}^z$) of the maximum response times of the $\text{NP}^{50}_{t_{\text{Pd}}}$ films. The exponent value of $z \approx 2.3$ agrees well with the diffusion limited case in metal hydride systems.[13] Note that VSR will go as length $L$ or radius $r$ in nanocubes and nanoparticles, which are traditionally used in power law fittings. In the Pd hemispherical caps on 500 nm PS beads VSR will be close to $t_{\text{Pd}}$ for small thicknesses (< 20 nm), but will begin to deviate for increasing thicknesses.

## S3. Optical properties of control samples

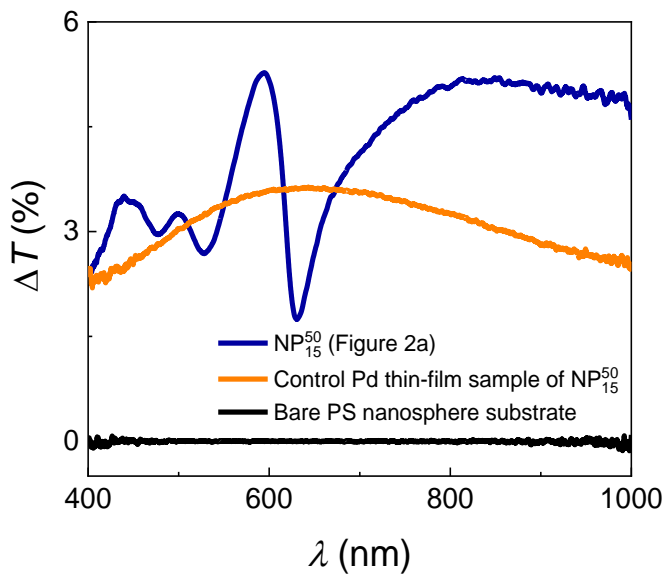

**Figure S14.** Optical transmission change $\Delta T(\lambda) = T_{1000\text{ mbar}} - T_{0\text{ mbar}}$ of PS nanosphere monolayer substrate, $\text{NP}_{15}^{50}$ sample, and corresponding control Pd thin-film sample ($t_{Pd}$ = 15 nm, θ = 50°).

The maximum optical change in the NP sample is much greater than that of the control Pd thin film sample (Figure S14). This can be attributed to diffraction within the hexagonal Pd patchy arrays, which does not occur in a thin film. Transmitted light at certain wavelengths is diffracted, partially localized inside the PS nanosphere causing the so-called partially-localized surface plasmon resonance (LSPR), and then is amplified through interactions with the Pd hemisphere cap.[14,15] This LSPR is suppressed when Pd turns to Pd hydrides, inducing significant redistribution of the local electric field.[14] This enhanced light-material interaction emerges in the far-field as local extrema on Δ$T$ spectra. To confirm this hypothesis, FDTD calculations were performed using NP morphology generated by a home-built MATLAB program,[1] and the calculated results are in excellent agreement with the experimental transmission spectra (Figures S15b-c). The FDTD calculated time-averaged intensity $|E/E_0|^2$ maps (square of ratio of electric fields E to the incident field $E_0$), extracted at the NPs cross-section at $\Delta T(\lambda)$ peaks, indicates enhanced localized electric field under the Pd patchy cap (Figures S15d–f), confirming the near- to far-field process.

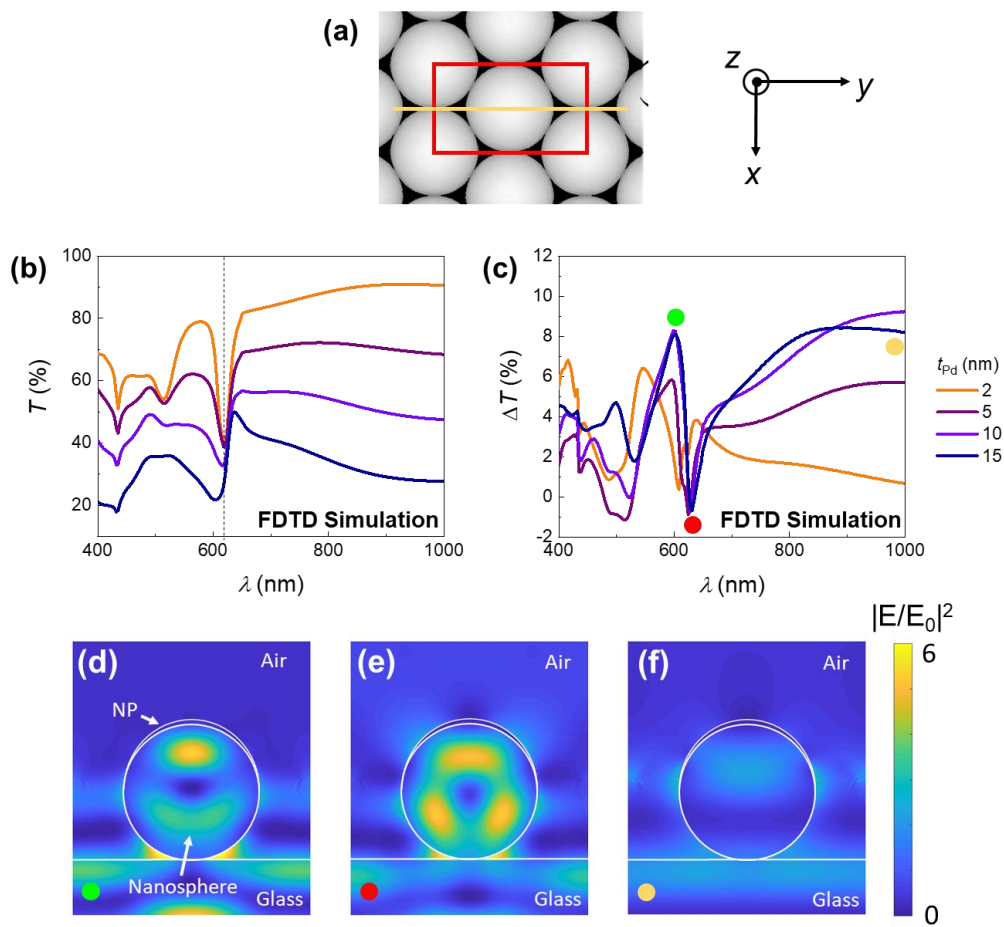

**Figure S15.** (a) A top-view of the hexagonal lattice of NP, the red box denotes the rectangular unit cell for FDTD calculations. (b) FDTD calculated optical transmission spectra $T(\lambda)$ of NP sample, when $P_{H_2}$ = 0 mbar. (c) FDTD calculated $\Delta T(\lambda) = T_{1000\ mbar} - T_{0\ mbar}$ spectra of NP samples. (d-f) Time-averaged intensity maps of the FDTD calculated local electric field at the cross-section plane denoted by the yellow line (Figure S15a). Circle dots with different colors (bottom left of the figures) indicate the wavelength that the map is extracted (Figure S15c). We observe an partial-localized enhanced electric field spot located under the Pd NP at $\Delta T(\lambda)$ peak (at green circle, Figure S15c), and a whispering-gallery modes inside nanosphere at $\Delta T(\lambda)$ dip (at red circle, Figure S15c). On the other hand, no enhanced electric field can be observed when it is out of resonance (at yellow circle, Figure S15c).

## S4. Plateau pressures extraction

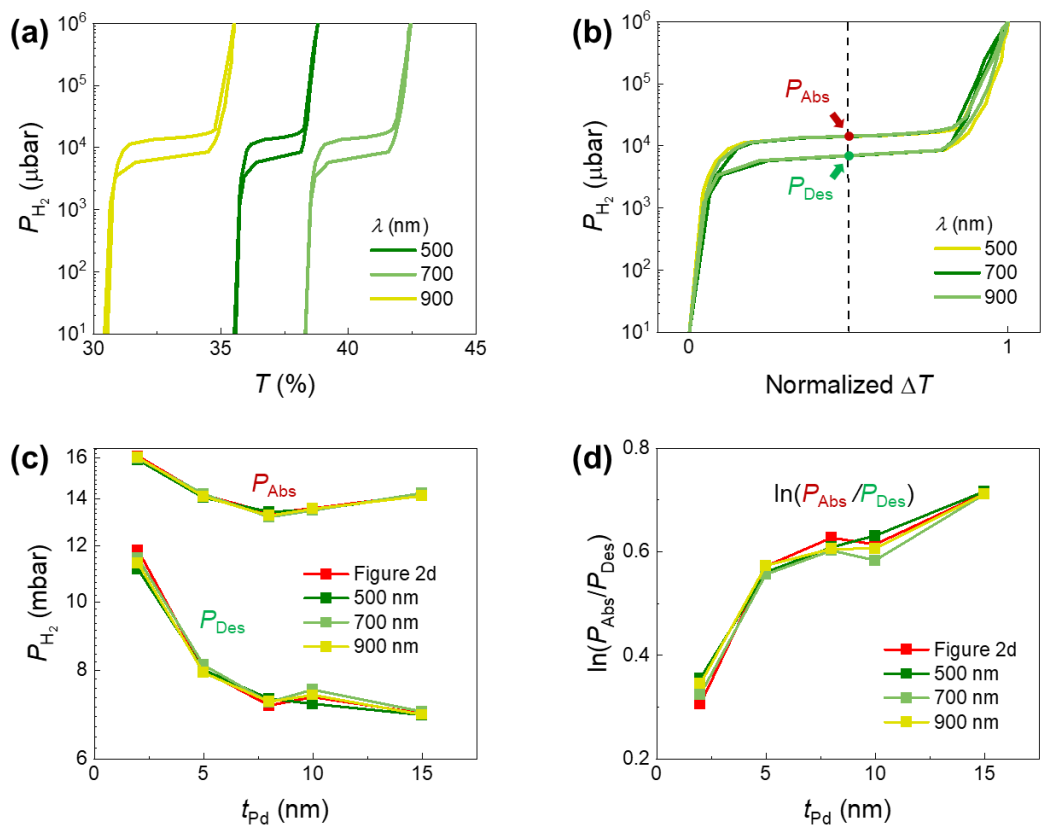

**Figure S16.** (a) Optical hydrogen sorption isotherm of representative Pd NP$_{15}^{50}$ sample, extracted at different wavelength positions and (b) normalized $\Delta T$ sorption isotherm at the corresponding wavelengths. Absorption and desorption plateau pressures ($P_{Abs}$ and $P_{Des}$, respectively) are extracted at normalized $\Delta T = 0.5$ (as denoted in Figure S16b). (c) $P_{Abs}$, $P_{Des}$, and (d) $\ln(P_{Abs}/P_{Des})$ extracted at different wavelength positions. We observe a slight variation of these values at different wavelengths; however, the general thickness-dependent trend is conserved.

## S5. Sensor accuracy calculations

The sensor accuracy ($A$) is calculated by the following equation:[16]

$$A = \frac{|log(P_{Abs}) - log(P_{Des})| \times 100}{|log(P_{Abs}) + log(P_{Des})|/2}, \quad (S20)$$

where $P_{Abs}$ and $P_{Des}$ are the pressures reading (in µbar) during hydrogen absorption and desorption, respectively.

## S6. The calculation of void coverage

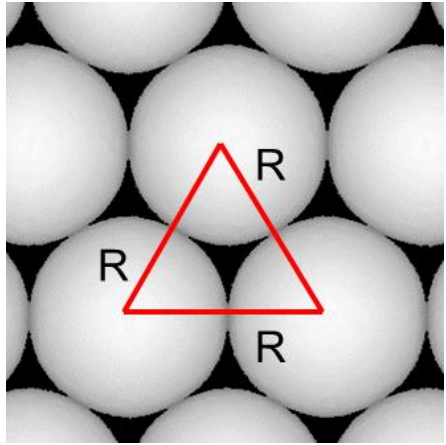

**Figure S17.** The calculation configuration for void coverage.

The void coverage percentage V (black area) between PS nanosphere (radius of R) can be calculated as:

$$V = (\frac{1}{2} \times 2R \times \frac{\sqrt{3}}{2}R - 3 \times \frac{1}{6} \times \pi R^2)/\frac{1}{2} \times 2R \times \frac{\sqrt{3}}{2}R = \frac{\sqrt{3} - \pi/2}{\sqrt{3}} \approx 0.093$$

## S7. Phase transition behaviors of NP sample with different θ

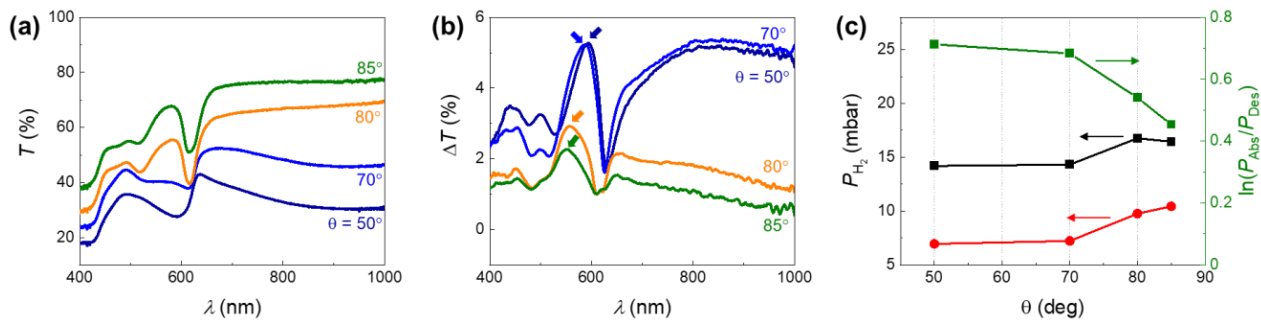

**Figure S18.** (a) Experimental optical transmission spectra $T(\lambda)$ (at $P_{H_2}$ = 0 mbar) and (b) optical transmission change $\Delta T(\lambda) = T_{1000\ mbar} - T_{0\ mbar}$ of NP with different vapor incident angles of θ. (c) Extracted plateau pressures for hydrogen absorption ($P_{Abs}$), desorption ($P_{Des}$), and $\ln(P_{Abs}/P_{Des})$ for different θ.

# S8. Optical properties of Pd, PdAg, PdAu, and PdCo composite NP upon hydrogenation and their sensing performances

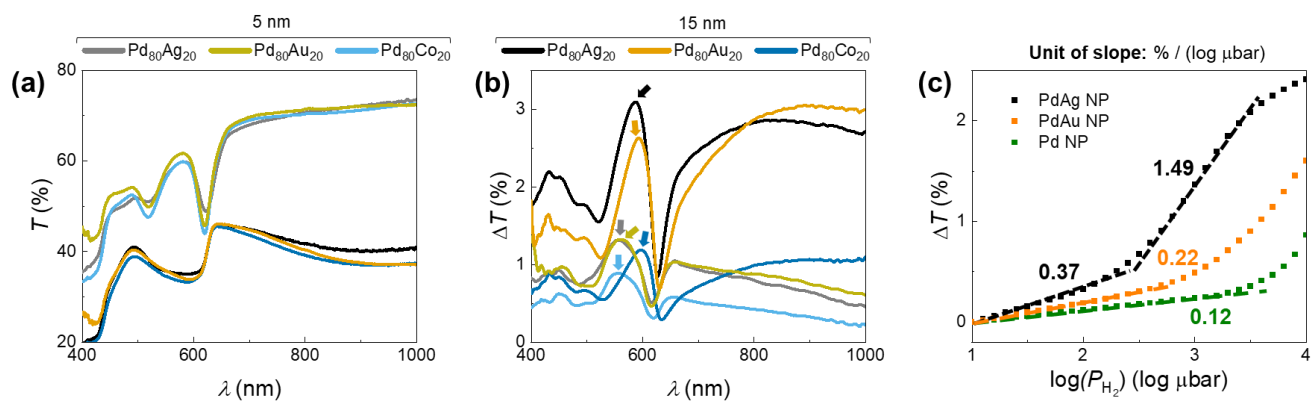

**Figure S19.** (a) Experimental optical transmission spectra $T(\lambda)$ and (b) optical transmission changes $\Delta T(\lambda) = T_{1000\ mbar} - T_{0\ mbar}$ of $Pd_{80}Ag_{20}$/$Pd_{80}Au_{20}$/$Pd_{80}Co_{20}$ $NP_{15}^{50}$ and $NP_{5}^{50}$ samples. Colored arrows denote the position of $\Delta T(\lambda)$ spectra peaks, where sorption isotherms are extracted (results presented in Figure 4b). (c) Extracted $\Delta T$ absorption isotherm of $Pd_{80}Ag_{20}$, $Pd_{80}Au_{20}$ and Pd $NP_{15}^{50}$ samples at spectra peak. The sensitivity numbers indicate the change in relative transmission magnitude (%) per 1 (log μbar) increase in $\log(P_{H_2})$.

In general, incorporating Ag or Au into NPs does not change the overall shape of $\Delta T(\lambda)$ (at $P_{H_2}$ = 1000 mbar), but it reduces the absolute value of $\Delta T(\lambda)$ ~50 % in comparison to that of pure Pd NP with the same deposited thickness (Figure S19). The drop of $\Delta T(\lambda)$ magnitude is more significant than the amount of Pd atoms replaced by Ag or Au (20%), which can be explained by the reduction in the limiting solubility of H in the PdAg or PdAu system. This reduction is due to a decreasing number of available electron states in the d-band of the Pd electronic structure, which is induced by the introduction of Ag/Au atoms.[17] In contrast, at very low pressures ($P_{H_2}$ < 1 mbar), the $Pd_{80}Ag_{20}$ and $Pd_{80}Au_{20}$ alloy NP sensor shows about 3- and 1.8-times enhancement in sensitivity in comparison to that of pure Pd NP (Figure S19c), respectively, which implies an increase of H solubility at low pressures arising from the interaction between an interstitial hydrogen and Ag/Au in the lattice.[17]

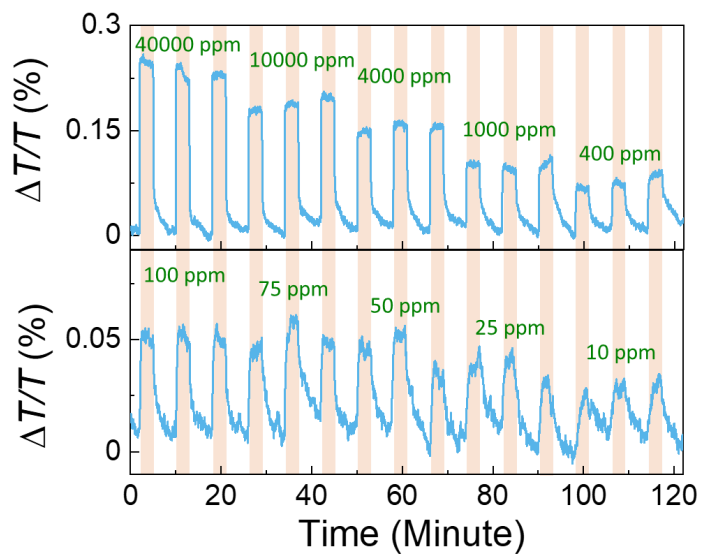

**Figure S20.** $\Delta T/T$ response of $Pd_{80}Co_{20}$ ($t$ = 5 nm) composite NPs (1.25 Hz of sampling frequency) with different hydrogen concentration ($C_{H_2}$), measured in flowing nitrogen (400 ml/min). Shaded areas denote the periods where the sensor is exposed to hydrogen.

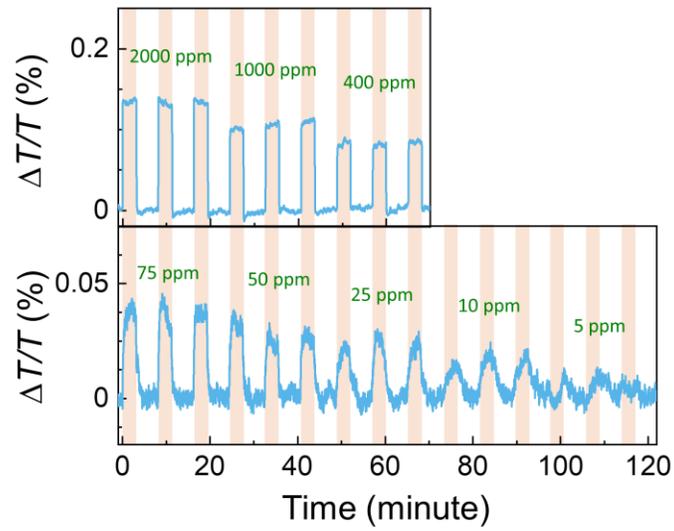

**Figure S21.** $\Delta T/T$ response of $Pd_{80}Co_{20}$ ($t$ = 5 nm) composite NPs (1.25 Hz of sampling frequency) with different hydrogen concentration ($C_{H_2}$), measured in flowing synthetic gas (400 ml/min). Shaded areas denote the periods where the sensor is exposed to hydrogen.

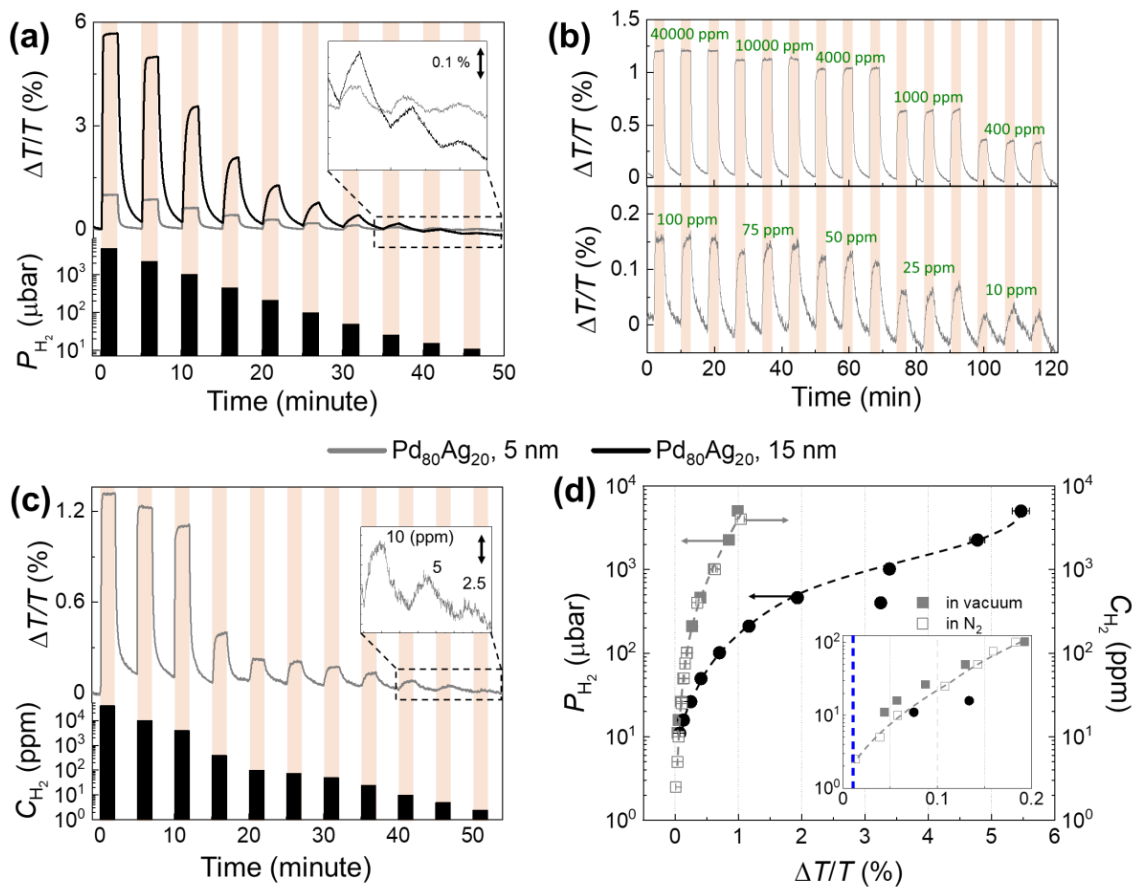

**Figure S22.** (a) $\Delta T/T$ response of $Pd_{80}Ag_{20}$ $NP_{15}^{50}$ and $NP_{5}^{50}$ sensors to stepwise decreasing hydrogen pressure in the 5000 – 11 μbar range, measured at 1.25 Hz sampling frequency in a vacuum chamber. (b)-(c) $\Delta T/T$ response of $Pd_{80}Ag_{20}$ $NP_{5}^{50}$ sensors with different hydrogen concentrations ($C_{H_2}$) of 4 % - 2.5 ppm, measured in flowing nitrogen (400 ml/min). Up-down arrows in the insets of (c) correspond to 0.03 %. Shaded areas denote the periods where the sensor is exposed to hydrogen. (d) Measured $\Delta T/T$ response as a function of hydrogen pressure/concentration derived from (a-c). The blue dashed line denotes the defined LOD at $3\sigma \approx 0.013$ % (see Section S9).

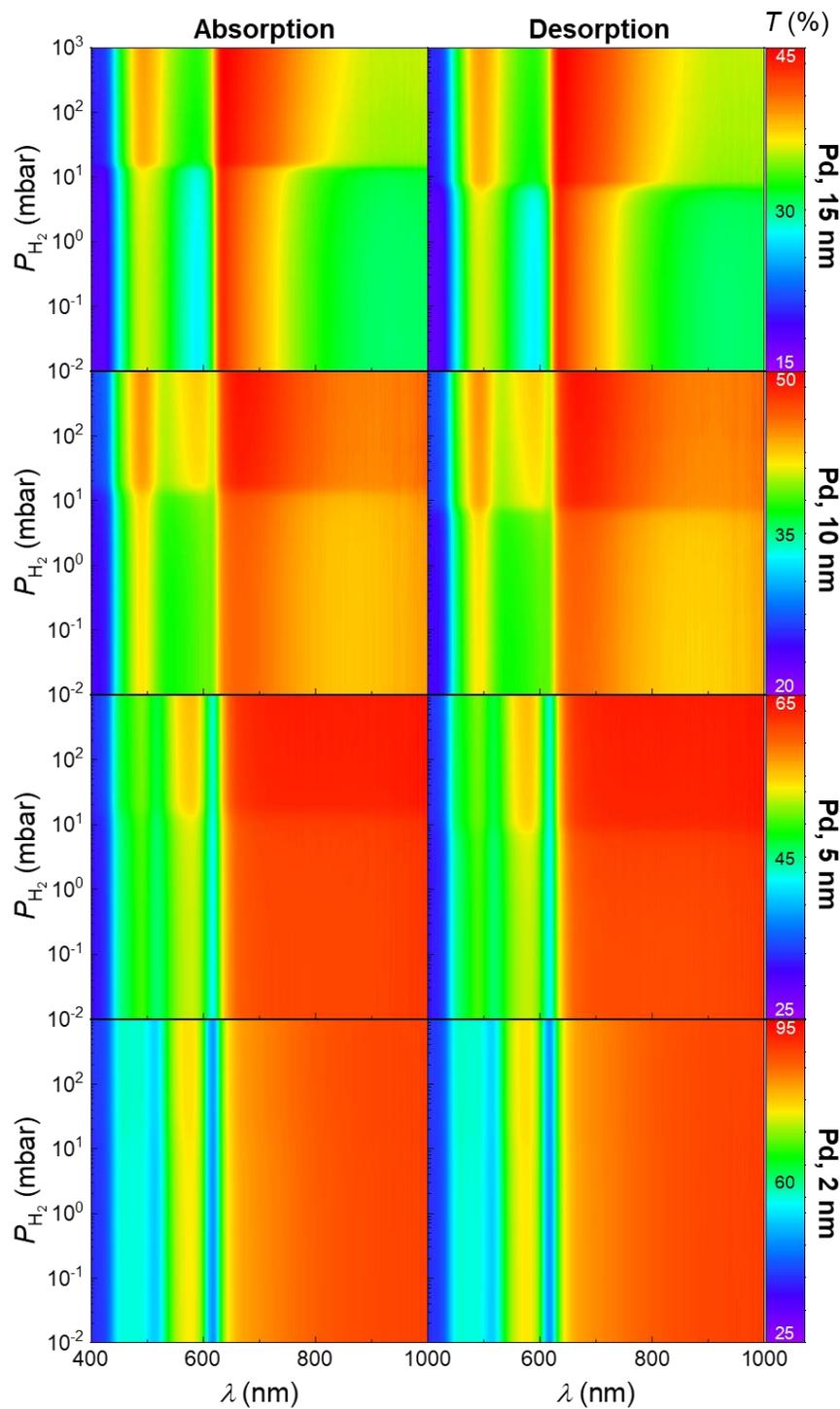

**Figure S23.** Spectra response of Pd NP$_t^{50}$ optical hydrogen sensor ($t$ = 2, 5, 10, and 15 nm) to an increasing (left column)/decreasing (right column) hydrogen pressure.

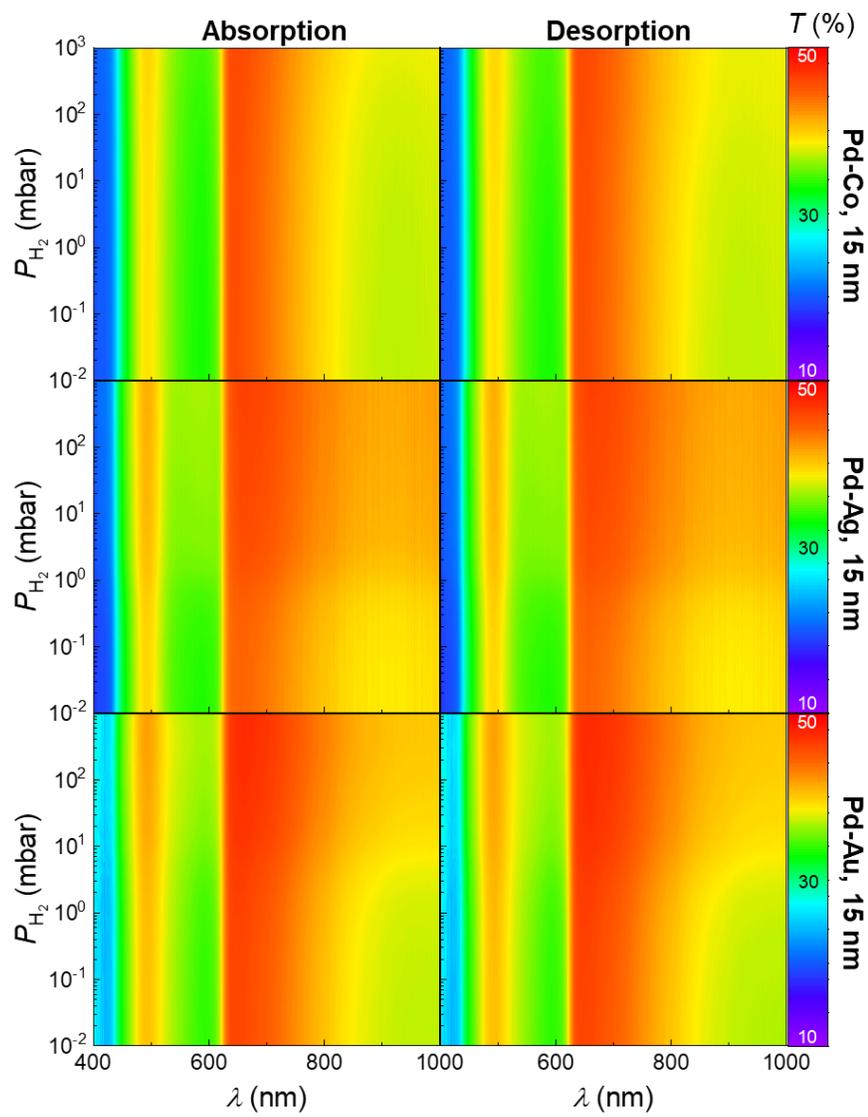

**Figure S24.** Spectra response of $NP_{15}^{50}$ optical hydrogen sensor (with different film composition) to an increasing (left column)/decreasing (right column) hydrogen pressure.

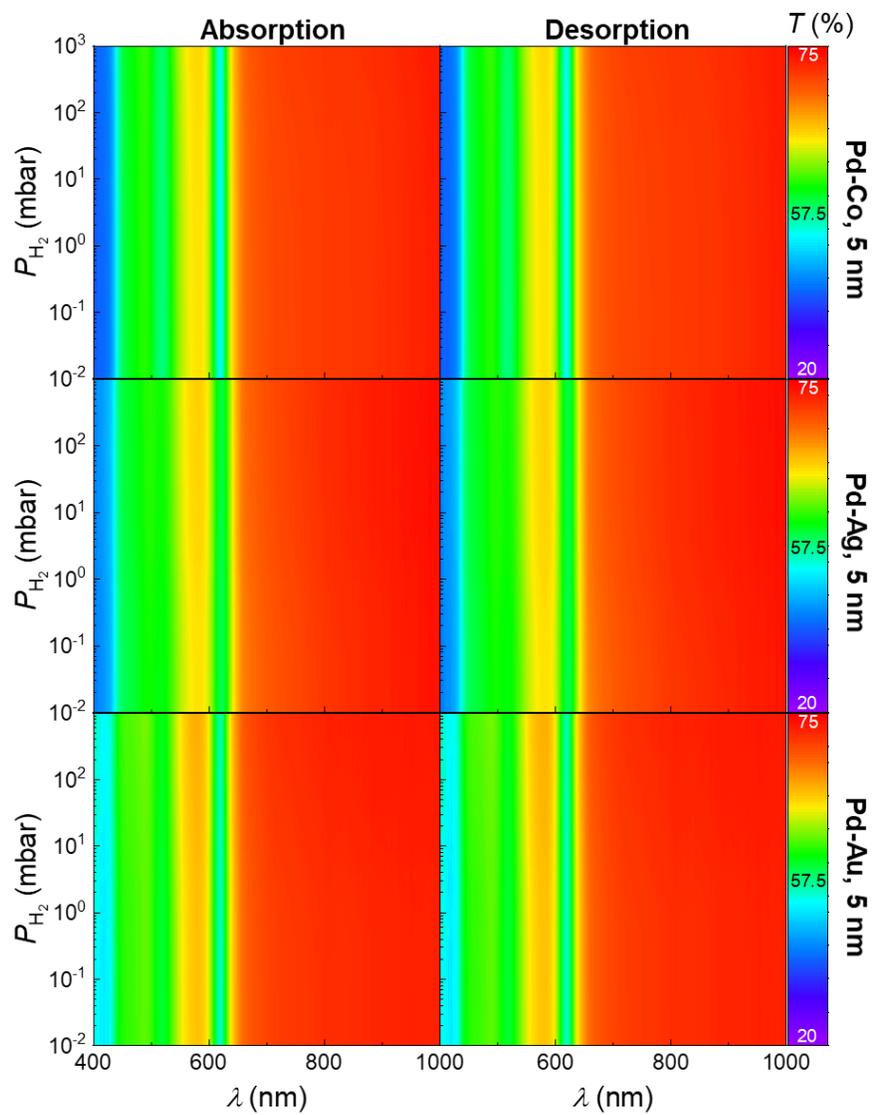

**Figure S25.** Spectra response of NP$_5^{50}$ optical hydrogen sensor (with different film composition) to an increasing (left column)/decreasing (right column) hydrogen pressure.

## S9. Noise evaluation

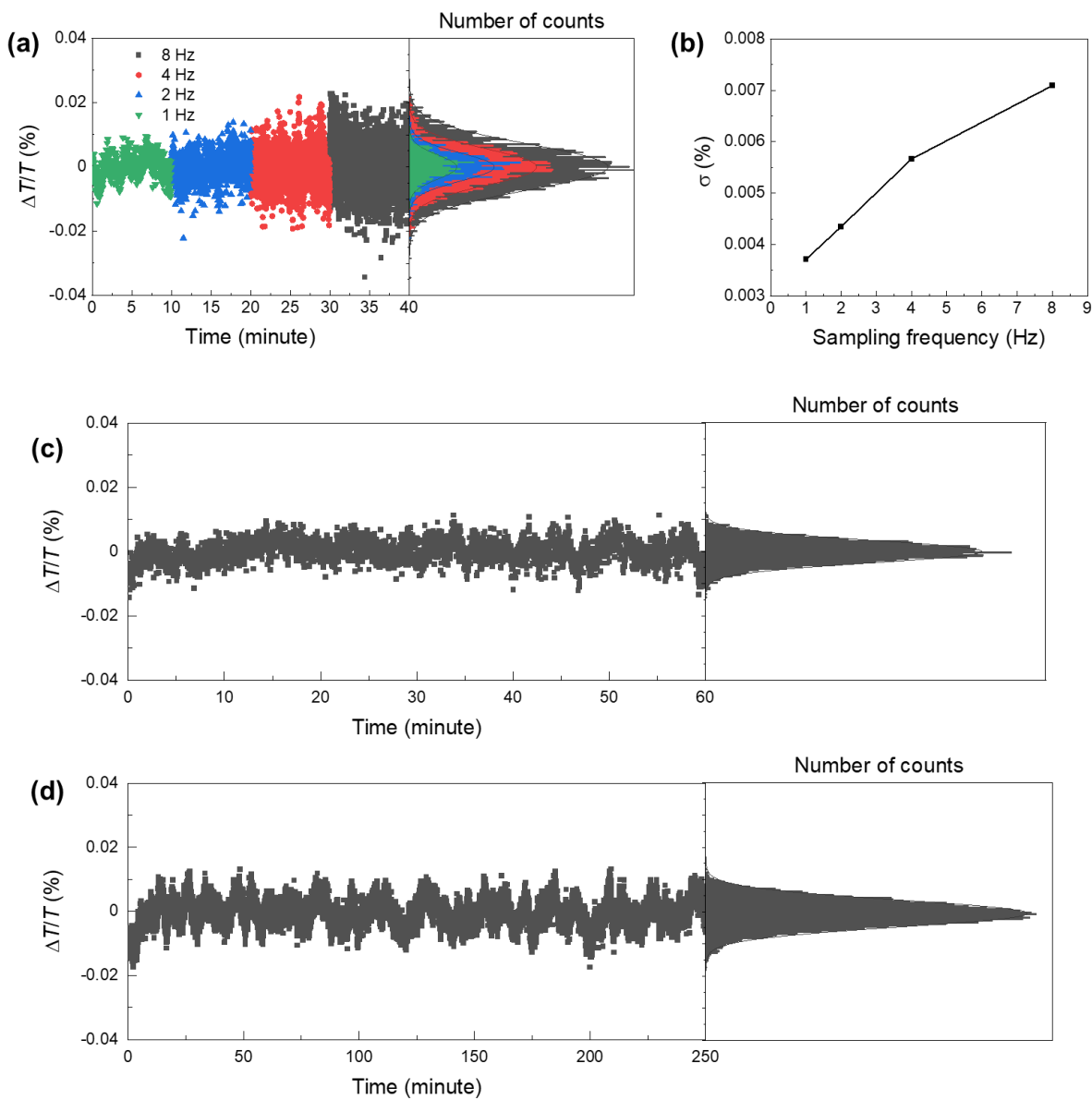

**Figure S26.** (a) Experimental signal noise with different sampling frequencies of PdCo NP$_5^{50}$ sample, measured at 0 mbar of H$_2$ (left) and histogram plot of signal intensity (right). The distribution of noise exhibits normal distribution with standard deviations of σ. (b) The experimental signal noise, σ, as a function of sampling frequency. At 1 Hz of sampling frequency, σ = 0.0037 %. (c) Experimental signal noise at 1 Hz of sampling frequency of PdCo NP$_5^{50}$ sample, measured in flowing 4 % hydrogen in nitrogen (400 ml/min) (left), and histogram plot of signal intensity (right). At 1 Hz of sampling

frequency, $\sigma$ = 0.0035 %. (d) Experimental signal noise at 1 Hz of sampling frequency of PdCo NP$_5^{50}$ sample, measured in flowing 2 % hydrogen in synthetic air (400 ml/min) (left), and histogram plot of signal intensity (right). At 1 Hz of sampling frequency, $\sigma$ = 0.0042 %.

# S10. PdCo $NP_5^{50}$ and PdCo $NP_5^{50}$/PMMA sensors stability

For a long-term stability assessment, we simply placed $Pd_{80}Co_{20}$ $NP_5^{50}$ sample in the ambient condition (temperature 22-23°C, humidity 35% RH) and characterize the response time and limit of detection (LOD) of the sensor in a weekly basis (week 1 to 6). These obtained results are then compared to a 10-month-old sample, in Figure S27.

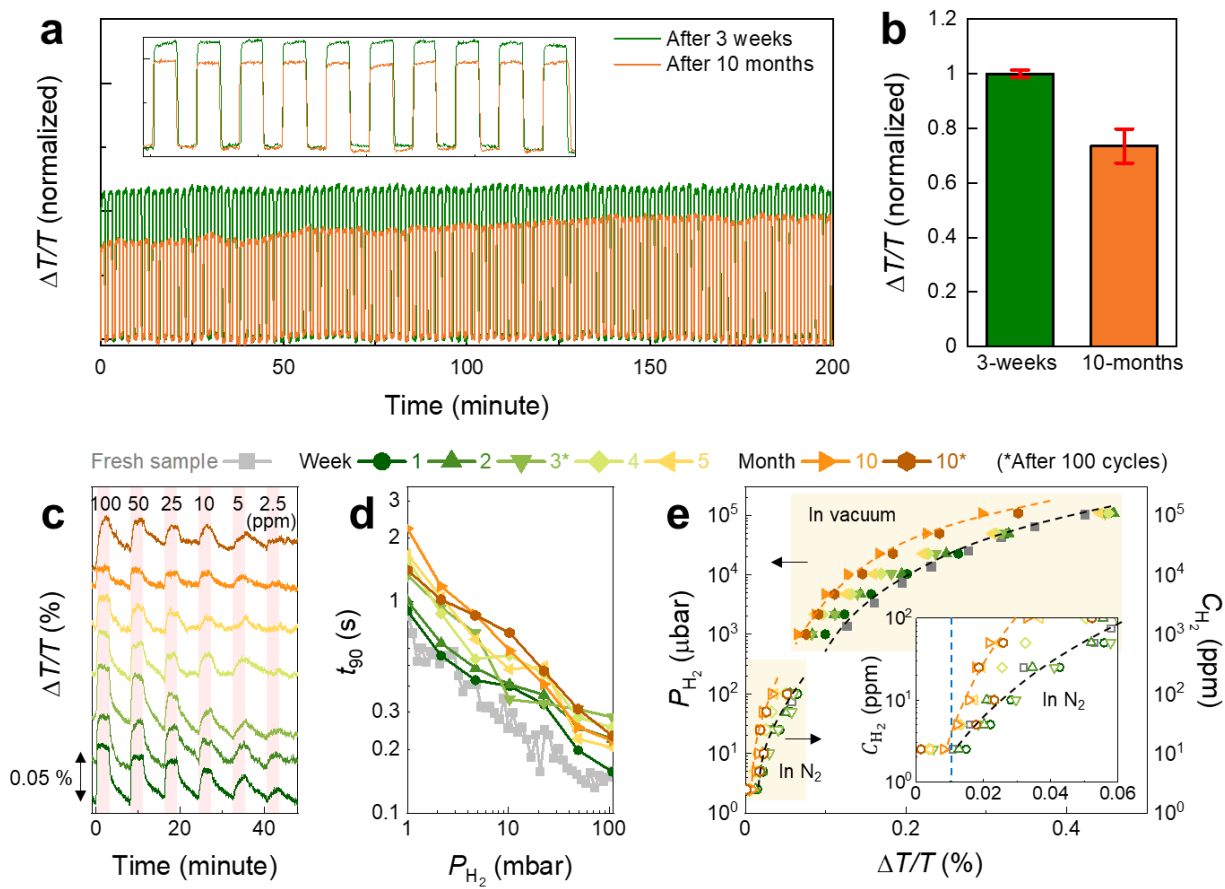

**Figure S27.** (a) $\Delta T/T$ response of $Pd_{80}Co_{20}$ $NP_5^{50}$ (3-weeks and 10-months old) upon 100 cycles (1/1 minute of loading/unloading) of 2% $H_2$ in synthetic gas (400 ml/min) and (b) $\Delta T/T$ response normalized to the response of a fresh sample, obtained in the same condition. The error bars denote the standard deviation from 100 cycles. (c) Long-term $\Delta T/T$ response of $Pd_{80}Co_{20}$ $NP_5^{50}$ with different hydrogen concentrations ($C_{H_2}$) of 100 – 2.5 ppm, measured in flowing nitrogen (400 ml/min). Shaded areas

denote the periods where the sensor is exposed to hydrogen. (d) Long-term response time of $Pd_{80}Co_{20}$ $NP_5^{50}$ with 1-100 mbar pure hydrogen pulse. (e) Measured $\Delta T/T$ response as a function of $P_{H_2}$ in vacuum/pure hydrogen (solid symbol) and $C_{H_2}$ in flowing nitrogen (400 ml/min) (half-up filled symbols). Inset: the blue dashed line denotes the defined LOD at $3\sigma \approx 0.011$ % ($\sigma = 0.0035$ %, is the noise of the acquired signal with $N_2$ carrier air, see Section S9).

The $Pd_{80}Co_{20}$ $NP_5^{50}$ sample shows a good stability without the sign of degradation, upon >300 of (de)hydrogenation cycles. The first 100 cycles (1/1 minute of loading/unloading with 2% $H_2$ in synthetic gas) of 3-weeks old and 10-months old sample are summarized in Figure S27a. While a noticeable reduction of sensor signal due to aging can be seen in 10-months old sample (Figure S27b), upon cycling, the signal is recovering back to that of a fresh sample.

The sign of the aging effect can be seen by the degraded performances of the sensor over the times. In particular, the response time $t_{90}$ (at $P_{H_2} = 1 - 100$ mbar) increases about ~2 times and ~3 times after a period of 4-weeks and 10-months, respectively (Figures S27c). In addition, the long-term $\Delta T/T$ responses show a significant reduction both in vacuum mode and flow mode (Figures S27d and e), just after 4-5 weeks in air. We ascribed the degradation of the sensor to a small trace amount of poison gases exist in the ambient air, (e.g. CO) (see the deactivation test with $CH_4$, $CO_2$, and CO in Figure 6). However, we note that the response time and LOD of the sample still are <2.5 s (at 1 mbar) and <10 ppm, respectively, over a >10-month period. The degradation of the $Pd_{80}Co_{20}$ $NP_5^{50}$ sample upon long-term storage in air can affect to the accuracy, sensitivity, and response time of the sensor. Hence, we look for a solution to mitigate this process. Inspired by the work of Nugroho et al.,[18] we coated the $Pd_{80}Co_{20}$ $NP_5^{50}$ sample with a ~50-nm layer of polymethyl methacrylate (PMMA) (by spin-coating of PMMA dissolved in acetone, more details can be found in Methods section), which has been demonstrated to effectively block the poisonous species. In this test, we store the PMMA-coated and uncoated samples in an identical condition and measure their sensing metrics.

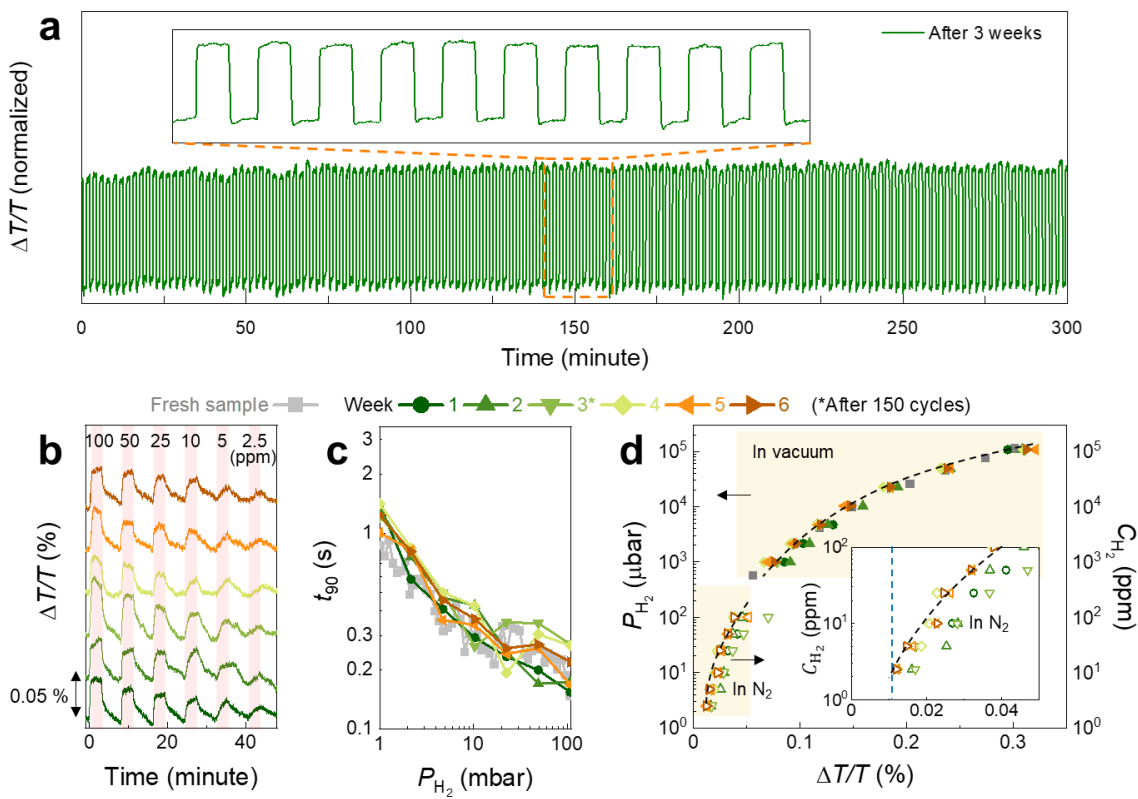

**Figure S28.** (a) $\Delta T/T$ response of $Pd_{80}Co_{20}$ $NP_5^{50}$/PMMA (3-weeks old) upon 150 cycles (1/1 minute of loading/unloading) of 2% $H_2$ in synthetic gas (400 ml/min). (b) Long-term $\Delta T/T$ response of $Pd_{80}Co_{20}$ $NP_5^{50}$/PMMA with different hydrogen concentrations ($C_{H_2}$) of 100 – 2.5 ppm, measured in flowing nitrogen (400 ml/min). Shaded areas denote the periods where the sensor is exposed to hydrogen. (c) Long-term response time of $Pd_{80}Co_{20}$ $NP_5^{50}$/PMMA with 1-100 mbar pure hydrogen pulse. (d) Measured $\Delta T/T$ response as a function of $P_{H_2}$ in vacuum/pure hydrogen (solid symbol) and with different $C_{H_2}$ in flowing $N_2$ (half-up filled symbols). Inset: the blue dashed line denotes the defined LOD at $3\sigma \approx 0.011$ % ($\sigma = 0.0035$ %, is the noise of the acquired signal with $N_2$ carrier air, see Section S9).

The sensing performances of $Pd_{80}Co_{20}$ $NP_5^{50}$/PMMA sensor are summarized in Figure S28. After storing in air for 6-weeks and underwent >200 cycles of (de)hydrogenation with 2% $H_2$, we observe very little variances in response time over the pressure range of 1-100 mbar ($t_{90}$ are <1.5 s (at 1 mbar)) and insignificant reduction of sensor signal upon exposure to pulses of very low $H_2$ concentration (100 – 2.5 ppm). Clearly, the degradation of the sensor performance is significantly slowed-down with a polymer coating layer.

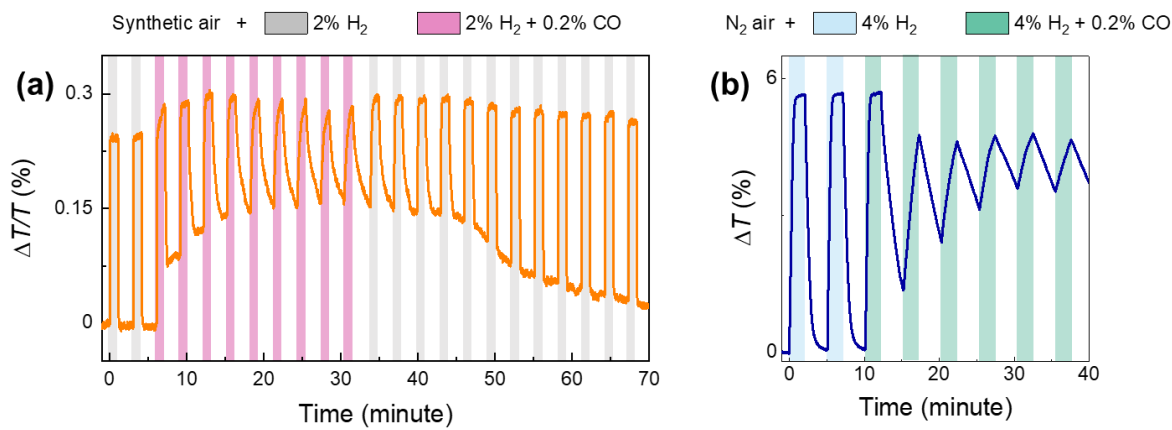

**Figure S29.** (a) Time-resolved $\Delta T/T$ response of $Pd_{80}Co_{20}$ $NP_5^{50}$ (to 2 pulses of 2% $H_2$ followed by 9 pulses of 2% $H_2$ + 0.2% CO, and 12 pulses of 2% $H_2$, with synthetic air as a carrier gas), shows that sensor signal and slow response/release time of a poisoned sensor can be recovered upon several (de)hydrogenation cycles. Note that the $\Delta T/T$ responses in this figure and Figure 6a are identical. (b) Time-resolved $\Delta T/T$ response of Pd $NP_{15}^{50}$ (to 2 pulses of 4% $H_2$ followed by 6 pulses of 4% $H_2$ + 0.2% CO, with nitrogen air as a carrier gas) for comparison purposes.

## S11. Sensing performances of a fresh PdCo $\text{NP}_5^{50}$/PMMA sensors

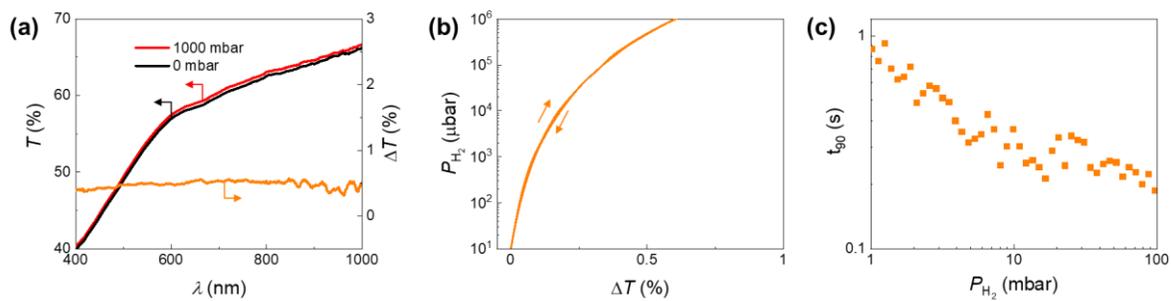

**Figure S30.** (a) Experimental optical transmission spectra $T(\lambda)$ at $P_{H_2}$ = 0 and 1000 mbar, and optical transmission changes $\Delta T(\lambda) = T_{1000\ mbar} - T_{0\ mbar}$ of PdCo $\text{NP}_5^{50}$/PMMA sensor. (b) Optical hydrogen sorption isotherm extracted at $\Delta T(\lambda)$ maxima. Arrows denote the sorption direction. (c) Response time of sensors with pulse of hydrogen pressure from 1 mbar to 100 mbar.

## S12. Sensing metrics of state-of-art optical hydrogen sensor (at room-temperature)

| Sensing platform | | $t_{90}$ (s) (@ 40 mbar) | $t_{90}$ (s) (@ 1 mbar) | LOD (ppm) | Hysteresis-free? | Ref. |
|---|---|---|---|---|---|---|
| $Pd_{80}Co_{20}$ NP | | ≤ 0.15 | 0.85 | 2.5 | Yes | This work |
| PdAu nano-particles @PTFE@PMMA | 190 × 25 (nm$^2$) | 0.3 | > 2 | 10<br>3 (by extrapolation) | Yes | 18 |
| | 100 × 25 (nm$^2$) | < 0.3 | 1 | n.a.*<br>< 1000 (estimated) | | |
| Pd nano-disk array | | > 10 | - | 50 | n.a. | 19 |
| Pd bilayer lattices | | ~ 900 | 55 | - | n.a. | 14 |
| PdAuCu nano-particles | | 0.4 | - | 5 | Yes | 20 |
| PdAu nanostructures | | 40 | - | - | Yes | 21 |
| Pd strip | | - | 20 | 10 | n.a. | 22 |
| PdY film | | 6 | - | 1000 | n.a. | 23 |
| Pd/SiO2/Au | | 3 | - | 5000 | n.a. | 24 |
| Pd/Au film | | 4.5 | - | - | n.a. | 25 |

**Table S1.** Sensing metrics of state-of-art optical hydrogen sensor (at room-temperature). *not addressed.